\documentclass[twocolumn]{aastex631}

\newcommand{\RN}[1]{\uppercase\expandafter{\romannumeral #1\relax}}

\usepackage{url}
\usepackage{threeparttable}
\shorttitle{New HVSs from LAMOST}
\shortauthors{Sun et al.}
\graphicspath{{./}{figures/}}
\begin{document}

\title{Systematic search for blue hyper-velocity stars from LAMOST survey}

\correspondingauthor{Yang Huang}
\email{huangyang@ucas.ac.cn}

\author[0000-0002-3935-2666]{Yongkang Sun}
\affiliation{Key Laboratory of Optical Astronomy, National Astronomical Observatories, Chinese Academy of Sciences, Beijing 100101, China}
\affiliation{School of Astronomy and Space Science, University of Chinese Academy of Sciences, Beijing 100049, China}

\author[0000-0003-3250-2876]{Yang Huang}
\affiliation{School of Astronomy and Space Science, University of Chinese Academy of Sciences, Beijing 100049, China}
\affiliation{Key Laboratory of Optical Astronomy, National Astronomical Observatories, Chinese Academy of Sciences, Beijing 100101, China}

\author[0000-0002-2874-2706]{Jifeng Liu}
\affiliation{Key Laboratory of Optical Astronomy, National Astronomical Observatories, Chinese Academy of Sciences, Beijing 100101, China}
\affiliation{School of Astronomy and Space Science, University of Chinese Academy of Sciences, Beijing 100049, China}

\author[0009-0002-4542-8046]{Haozhu Fu}
\affiliation{Department of Astronomy, Peking University, Beijing, 100871, China}
\affiliation{Kavli Institute for Astronomy and Astrophysics, Peking University, Beijing, 100871, China}

\author[0000-0002-7727-1699]{Huawei Zhang}
\affiliation{Department of Astronomy, Peking University, Beijing, 100871, China}
\affiliation{Kavli Institute for Astronomy and Astrophysics, Peking University, Beijing, 100871, China}

\author[0000-0001-7607-2666]{Yinbi Li}
\affiliation{Key Laboratory of Optical Astronomy, National Astronomical Observatories, Chinese Academy of Sciences, Beijing 100101, China}
\affiliation{School of Astronomy and Space Science, University of Chinese Academy of Sciences, Beijing 100049, China}

\author[0000-0002-3954-617X]{Cuihua Du}
\affiliation{School of Astronomy and Space Science, University of Chinese Academy of Sciences, Beijing 100049, China}

\author[0000-0002-0349-7839]{Jianrong Shi}
\affiliation{Key Laboratory of Optical Astronomy, National Astronomical Observatories, Chinese Academy of Sciences, Beijing 100101, China}
\affiliation{School of Astronomy and Space Science, University of Chinese Academy of Sciences, Beijing 100049, China}

\author[0000-0001-8011-8401]{Xiao Kong}
\affiliation{Key Laboratory of Optical Astronomy, National Astronomical Observatories, Chinese Academy of Sciences, Beijing 100101, China}




\begin{abstract}
Hypervelocity stars (HVSs) represent a unique class of objects capable of escaping the gravitational pull of the Milky Way due to extreme acceleration events, such as close encounters with the supermassive black hole at the Galactic center (GC), supernova explosions in binary systems, or multi-body dynamical interactions.
Finding and studying HVSs are crucial to exploring these ejection mechanisms, characterizing central black holes, probing the GC environment, and revealing the distribution of dark matter in our galaxy. The Large Sky Area Multi-Object Fiber Spectroscopic Telescope (LAMOST) spectroscopic surveys have so far identified four B-type unbound HVSs. To expand this sample with the second-phase LAMOST survey that started in 2018, we conducted a systematic search for early-type HVSs using the LAMOST Data Release 10. We identified 125 early-type high-velocity candidates with total velocities exceeding 300 km\,s$^{-1}$.
Among them, we report ten new unbound B- and A-type hypervelocity star (HVS) candidates (designated LAMOST-HVS5 through LAMOST-HVS14), tripling the number of unbound HVSs previously identified by LAMOST.
Kinematic analyses suggest that these newly discovered HVS candidates likely originated either from the Galactic Center or via dynamical interactions.
Future high-resolution follow-up observations promise to refine the stellar parameters, distances, and elemental abundances of these candidates, thereby providing deeper insights into their origins and broadening their potential applications across astrophysics.
\end{abstract}



\section{Introduction} \label{sec:intro}
Most stars in our Galaxy are bound by its gravitational potential. Hypervelocity stars (HVSs) are a rare group with exceptionally high velocities (as high as few 1000 km\,s$^{-1}$) relative to the Galactic center (GC), allowing them to escape the Galaxy. The most likely origin of HVSs is their ejection from the GC due to three-body interactions involving the central supermassive black hole(s) \citep[SMBH; e.g.,][]{hills1988hyper,yu2003ejection,ginsburg2007hypervelocity,brown2015hypervelocity}. Other possible origins include runaway stars ejected from the Galactic disk via supernova explosions in massive binary systems \citep[e.g.,][]{blaauw1961origin,2000ApJ...544..437P,perets2012properties}, Type Ia supernovae explosions \citep[SNe Ia;][]{wang2009companion,shen2018ApJ}, or dynamical interactions in young star clusters \citep[e.g.,][]{gvaramadze2009origin}. Additionally, HVSs can also be ejected from the Galactic sub-systems, such as Large Magellanic Cloud \citep[LMC;][]{boubert2016dipole,boubert2017hypervelocity}, the Sagittarius Dwarf Spheroidal galaxy \citep{huang2021ApJ,Li_Du_Ma_Shi_Newberg_Piao_2022,2023Li_AJ}, and the Galactic globular clusters \citep{10.1093/nsr/nwae347}.

Investigating the origin of HVSs is both important and intriguing (see a review in \citealt{2023Li_AJ}).  HVSs are closely linked to the environment of the GC and the central SMBH(s), providing insights into the nature of the SMBH(s) \citep{yu2003ejection} and the binary population at the GC \citep{bromley2006ApJ}. Furthermore, the trajectories of HVSs can probe the dark matter halo potential of our Galaxy \citep{gnedin2005ApJ,yu2007MNRAS,perets2009ApJ}. Discovering more HVSs is essential to achieve these goals.

The majority of known HVSs are massive early-type stars found in the Galactic halo. The first HVS was serendipitously discovered by \citet{brown2005discovery}, identified as a main-sequence B-type star. 
The second B-type HVS was discovered by \citet{edelmann2005he}, which is believed to have an LMC origin \citep{erkal2019hypervelocity}. 
The following dedicated surveys have led to the discovery of more than twenty B-type HVSs in total \citep{brown2006successful,brown2006hypervelocity,brown2007hyper,brown2007hypervelocity,brown_2009,brown_2009b,brown2012mmt,brown2014mmt, zheng2014,huang2017,li2018}.
Meanwhile, several promising candidate HVSs were identified by other serendipitous and systematic efforts \cite[e.g.,][]{tillich2009sdss, przybilla2008hd, irrgang2010nature, li2012ApJ, li2015RAA, du2019ApJS,2023Li_AJ}.
Most recently, \citet{2020MNRAS.491.2465K} discovered an A-type hypervelocity star (S5-HVS1) with a total velocity of $1755 \pm 50$ km~s$^{-1}$. Its backward-integrated trajectory 
point unambiguously to the GC, strongly indicating a GC origin.

The other types of HVSs are hot subdwarfs and white dwarfs (WDs), which are related to thermonuclear supernovae explosions. In the double-degenerate (DD) scenario (WD + WD binary), after the accretor's detonation, the lower-mass donor will be released with a velocity of 1000 - 2500 km s$^{-1}$, resulting in a hyper-velocity hydrogen-free WD \citep{shen2018ApJ...865...15S, el2023fastest}. This is also termed as the `D6 scenario' \citep{shen2018ApJ}. 
For single-degenerate (SD) scenario, the surviving non-degenerate donor star may be found with a high velocity.
This is likely the case for the first runaway hot subdwarf US 708 \citep{geier2015Sci,wang2009companion,wang2013producing}, which is a high-temperature helium-rich subluminous O-type star \citep{hirsch2005us}. It is also proposed that the accreting near-Chandrasekhar WDs in the SD systems may not be fully destroyed \citep[leading to SNe Iax;][]{jordan2012ApJ} to leave a low-mass remnant with a high peculiar velocity of $\sim 600$ km s$^{-1}$. The first such partially burnt remnant is LP 40-365 \citep[GD 492;][]{vennes2017Sci,raddi2018MNRAS}. So far, over 30 hyper-velocity WD candidates have been discovered \citep{shen2018ApJ...865...15S,el2023fastest,raddi2019MNRAS,el2023fastest,igoshev2023MNRAS}.

The discovery of distant HVSs remains challenging due to poor distance measurements, even in the {\it Gaia} era, which limits accurate determinations of their tangential velocities. However, their radial velocities can be precisely measured through large spectroscopic surveys.
The Large Sky Area Multi-Object Fiber Spectroscopic Telescope (LAMOST) is a powerful tool for searching for HVSs.
This is particularly true for early-type stars. Since early-type stars are significantly less numerous than late-type stars, LAMOST can target and survey them effectively and quickly.
 Until 2018, four early-type HVSs (LAMOST HVS 1-4, all B-type stars) have been discovered \citep{zheng2014,huang2017,li2018}. However, the search for HVSs from the LAMOST survey has stalled for years. Since 2018, LAMOST has begun the Phase II survey, and LAMOST data release 10 (DR10) published more than 11,000,000 low-resolution spectra in 2023, providing new opportunities to find more early-type HVSs. Here, after a long hiatus, we carry out a new systematic search for blue HVSs from LAMOST DR10, aiming to discover new classical massive early-type HVSs, as well as runaway supernova-related HVSs, which are all blue in colors.

The paper is organized as follows. In Section 2, we describe the sample selection HVS searching from LAMOST data. We present the results and discuss the properties of individual objects and their possible origin in Section 3. Finally, we summarize in Section 4.

\section{Data and Target Selection}

\subsection{LAMOST spectra}\label{sec:LAMOST}
The LAMOST is a four-meter aperture reflecting Schmidt telescope designed for efficiently obtaining spectra for both Galactic and extra-galactic objects in the optical band \citep{cui2012large,zhao2012lamost}. We utilize the LAMOST low-resolution spectra (LRS) that cover the wavelength of 3700-9000 \AA, with a spectral resolving power of about 1800 for finding HVS candidates. We use the values of $z$ (redshift) from the LAMOST LRS General Catalog to obtain the line-of-sight velocities ($v_{\rm{los}}$) of the stars.

\subsection{Gaia data}\label{sec:gaia}
The {\it Gaia} data release 3 \citep[DR3;][]{prusti2016gaia,gaiadr3} presented astrometric and photometric information for all-sky stars with {\it G} mag down to 21. We used the {\it Gaia} DR3 photometry data and early DR3 geometric distances {\it rgeo} \citep{bailer2021estimating} to compute the {\it Gaia} absolute magnitude $M_{G}$ and dereddened color $(BP-RP)_{\rm0}$. Based on this, we can then generate the {\it Gaia} color-magnitude diagram (CMD). The {\it Gaia} DR3 proper motion measurements are used for calculating the tangential velocities.

\subsection{Target Selection}\label{sec:target}
For finding early-type HVS and hyper-velocity WD/subdwarf candidates, we first select blue objects from the {\it Gaia} CMD for all LAMOST DR10 sources. Then we select out stars with high rest-frame radial velocities relative to the Galactic center ($v_{\rm{rf}}$), or with potential high Galactocentric total velocities ($v_{\rm{tot}}$) using \textit{rgeo} distances. Distances with large uncertainties will be redetermined using spectral indices or types.

\begin{figure}
    \includegraphics[width=1.0\columnwidth]{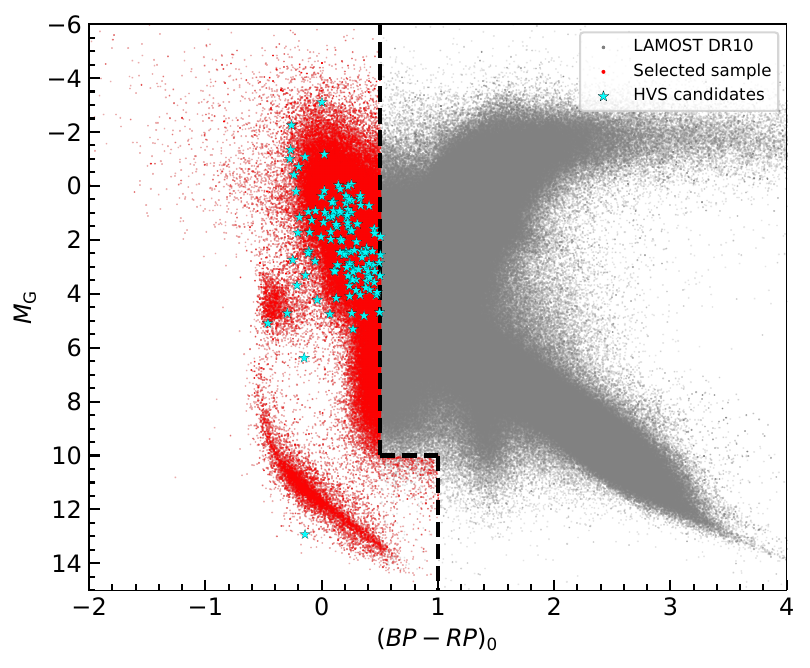}
	\caption{Gaia CMD for all LAMOST DR10 LRS sources (grey). The bluer part with SNR cut (red) represents stars (625,644 LAMOST LRS spectra in total) used to search for blue HVSs. The final HVS candidates containing 136 spectra of 124 objects are indicated as cyan stars.}
	\label{fig:cmd}
\end{figure}

For the first step, we present the {\it Gaia} CMD selection. The {\it Gaia} CMD for LAMOST sources is calculated with $E(B-V)$ from the 3D dust map provided by \citet{green2019ApJ} and reddening extinction coefficients from \citet{zhang2023ApJS}. We perform a color cut using the following criteria: $(BP-RP)_0 < 0.5$ or $((BP-RP)_0<1 $ and $ M_{G} > 10)$.
This cut enables early-type stars down to late-A-type stars and most sources at the white dwarf sequence to be retained. The color $(BP-RP)_0 < 0.5$ also guarantees that the potential D6 stars can be left in the selected region \citep{el2023fastest}. The previous work for finding HVSs based on LAMOST mainly focuses on late-type (FGK) stars \cite{2021ApJS..252....3L, 2023Li_AJ}. The loosed color cut when $M_{G} > 10$ enables the main WD cooling track to be included. Meanwhile, we exclude spectra with a signal-to-noise ratio (SNR) lower than 0.4 in {\it u} band and 1 in the {\it g} and {\it r} bands to avoid very-low quality spectra. The remaining $\sim 420,000$ objects (625,644 spectra) are plotted as red dots in Figure~\ref{fig:cmd}.
Such a loose SNR cut is applied to include as many candidate HVSs as possible, given that some LAMOST spectra have significantly underestimated SNRs. A thorough visual inspection will be conducted later to ensure all the spectra are of high quality.

We then employ two parallel criteria to select high-velocity candidates using either $v_{\rm{rf}}$ or $v_{\rm{tot}}$. First, we filter targets with $\left| v_{\rm{rf} } \right| > 300$ km s$^{-1}$. Since the reliability of $v_{\rm{rf}}$ depends solely on $v_{\rm{los}}$, genuine high-velocity stars can be identified based purely on LAMOST redshift measurements, following a visual inspection. We calculate the Galactocentric radial velocities for the targets using the following equation:
\begin{equation}
	v_{\rm{rf}} = v_{\rm{los}} + U_{\odot} {\rm{cos}} (b) {\rm{cos}} (l) + V_{\phi,\odot} {\rm{cos}} (b) {\rm{sin}} (l) + W_{\odot} {\rm{sin}}(b),
	\label{eq:vrf}
\end{equation}
where $l$ and $b$ are the Galactic longitude and latitude, receptively. 
$U_{\odot}=11.69$ km s$^{-1}$ and $W_{\odot}=7.67$ km s$^{-1}$ are the solar peculiar motion taken from \citet{wang2021MNRAS}. $V_{\phi,\odot}$ is calculated to be 252.23 km s$^{-1}$ from the proper motion ($-6.379$ mas~yr$^{-1}$ in Galactic longitude) of the Sgr A* \citep{reid2004ApJ} and the Galactocentric distance $R_0 = 8.34$ kpc of the Sun \citep{reid2014ApJ}. We also try other choices of solar peculiar motion and circular speed of the Sun (e.g.; \citealt{2010MNRAS.403.1829S,2015MNRAS.449..162H,2016MNRAS.463.2623H,2023ApJ...946...73Z}) and the whole results hold very well.

Second, we select targets with $v_{\rm{tot}} > 500$ km s$^{-1}$. The high threshold for $v_{\rm{tot}}$ can reduce the source number for further eye examination of LAMOST redshift measurement and the re-determination of unreliable distances. This approach comes at the cost of missing some objects with $v_{\rm{tot}}$ between 300 and 500 km s$^{-1}$, which may have low $v_{\rm{rf}}$ but relatively high tangential velocities. However, our primary goal is to identify distant HVSs with high $v_{\rm rf}$, as these velocities are more precisely measured than tangential velocities.

In order to calculate $v_{\rm{tot}}$, we first derive the three-dimensional (3D) space velocities ($U$, $V$, $W$) in Cartesian system centered on the solar position.
The space velocities are calculated with the {\it Gaia} DR3 positions and proper motions, the \textit{rgeo} distances \citep{bailer2021estimating} and the LAMOST line-of-sight velocities using the \textsc{Python} package \textsc{Pyastronomy} \citep{pya}, with a correction for the solar peculiar motion using ($U_\odot$,$V_\odot$,$W_\odot$) = (11.69, 10.16, 7.67) km~s$^{-1}$ \citep{wang2021MNRAS}.
The $v_{\rm{tot}}$ is then calculated from the following equation:

\begin{equation}
	v_{\rm{tot}} = \sqrt{{U}^2
		+(V+V_{c}~(R_{0}))^2
		+{W}^2},
	\label{eq:vtot}
\end{equation}
where $U$, $V$, and $W$ are the velocity component positive toward the Galactic center, in the direction of Galactic rotation, and toward the north Galactic pole, respectively. The circular rotation velocity at the solar position $V_{c}$~$(R_{0})$ is adopted to be 230 km s$^{-1}$ \citep{2023ApJ...946...73Z}

We further apply cuts $P_{\rm QSO} < 0.5$ and $P_{\rm Gal} < 0.5$ (from the {\it Gaia} DR3 catalog) to exclude potential extragalactic targets from the selected candidate HVSs.
In total, about 1000 high-velocity candidates are left by the two selection criteria: $\left| v_{\rm{rf} } \right| > 300$ km s$^{-1}$ or $v_{\rm{tot}} > 500$ km s$^{-1}$.
Then the spectral lines of these candidates are visually checked to discard targets with incorrect $v_{\rm{los}}$ measurements reported by the LAMOST pipeline. We also exclude stars with large $v_{\rm{los}}$ variations (probably due to binary orbital motion) if they have multiple spectroscopic observations. 
We also check the variability of light curve through the Zwicky Transient Facility \citep[ZTF;][]{2019PASP..131a8002B} data to eliminate eclipsing binaries and mark stars with periodic variability.

Finally, 136 spectra of 124 individual objects are selected as high-velocity candidates (star symbols shown in Figure~\ref{fig:cmd}). We find that five known HVSs, LAMOST-HVS1, 3 and 4, the bound HVS SDSS J081828.07+570922.1 discovered by \citet{brown2007hyper} and the hyper-runaway star HIP 60350 \citep{1998A&A...339..782M,irrgang2010nature} are recovered in our sample. 
We note that LAMOST-HVS2 was discovered using LAMOST test data, which is not included in the official release. As a result, it was not recovered in our search.

We summarize the selection result in Table~\ref{tab:selection}. We consider the parallax distances \textit{rgeo} to be unreliable if the \textit{rgeo} is larger than 6 kpc or the relative parallax error is larger than 0.2 (\textit{RPlx} $<$ 5). For objects with $\left| v_{\rm{rf} } \right| > 300$ km s$^{-1}$, they are truly high-velocity stars because the $v_{\rm{rf}}$ is independent of distances. However, the $v_{\rm tot}$ we used here to filter candidates (the $v_{\rm{tot}}$ in Table~\ref{tab:selection}) will not be reliable if the \textit{rgeo} distance is unreliable. Eight objects with $\left| v_{\rm{rf} } \right| < 300$ km s$^{-1}$ but $\left| v_{\rm{tot}} \right| > 500$ km s$^{-1}$ are potential HVS candidates, whose total velocities are required to be re-estimated. For stars with unreliable parallax distances, we will derive the photometric distances using the {\it Gaia} absolute magnitude $M_{G}$ from the \textsc{MKCLASS} method or from the Balmer line index of A-type stars. The photometric distances will confirm whether that eight potential HVS candidates have truly high velocities. Furthermore, the photometric distances for stars with $\left| v_{\rm{rf} } \right| > 300$ km s$^{-1}$ will help to determine whether they are unbound HVS candidates that exceed the Galactic escape velocities.

\begin{table*}
	\footnotesize
	%\scriptsize
	\centering
	\caption{Selection result of 124 high-velocity star candidates}
	\label{tab:selection}
	\begin{tabular}{ccccc} 
 \hline
 \hline
& with reliable {\tt rgeo}& without reliable {\tt rgeo}& Total number\\
 \hline
$v_{\rm{rf}} > 300$ km s$^{-1}$ & 38 objects (44 spectra) & 78 objects (84 spectra) & 116 objects (128 spectra)\\
(truly high-velocity stars)& & &\\
 \hline
$v_{\rm{rf}} < 300$ km s$^{-1}$
but $v_{\rm{tot}} > 500$ km s$^{-1}$& 0 objects & 8 objects (8 spectra) & 8 objects (8 spectra)\\ 
(potential HVS candidates)&&&(Exmained with photometric distances)\\
 \hline
Total number & 38 objects (44 spectra) & 86 objects (92 spectra) & 124 objects (136 spectra)\\
 \hline
    \end{tabular}
 \end{table*}

\subsection{Photometric distance} \label{sec:ld}

The \textsc{MKCALSS} code \citep{2014AJ....147...80G} is capable of deriving the spectral and luminosity class in the MK system \citep{1943assw.book.....M} for early-type stars. We apply the modified version \citep{2019ApJ...884L...7H} based on the original \textsc{MKCALSS} code to our HVS candidates to derive the best-fit MK classification of each LAMOST spectrum. For main-sequence (MS) stars, we utilize the table provided by Mamajek\footnote{\url{https://www.pas.rochester.edu/~emamajek/EEM_dwarf_UBVIJHK_colors_Teff.txt}} to obtain the $G$-band absolute magnitude $M_{G}$. The $1\sigma$ uncertainty in $M_{G}$ is taken as half the difference in $M_{G}$ between two neighboring spectral types.

For stars with a luminosity class lower than V, as classified by MKCLASS, there is no fixed absolute magnitude like that of main-sequence stars. 
Luickly, the intensity of Balmer lines has been shown to serve as an indicator of absolute magnitude for A-type stars \citep{1949AJ.....54..193P}. We thus select a low-extinction sample from the LAMOST LRS Line-Index Catalog of A-type Stars\footnote{\url{http://www.lamost.org/dr10/v1.0/catalogue}} and then examine the relation between the line indexes of various Balmer lines and the $G$-band absolute magnitudes $M_{G}$. We find that the relationship using the H$\delta$ line indices with 48~\AA~ band widths produced relatively small scatter in the residuals. Therefore, we adopted the H$\delta$ 48~\AA~ line indices from the catalog to infer $M_{G}$ (see Appendix~\ref{sec:appendix} for details). Finally, we apply a two-segment linear fit, as described by the following equations:
\begin{equation}
%\begin{aligned}
M_{G(A-type)} = -0.429\times \rm{Line Index48_{H\delta}}+5.943 
\end{equation}
if $\rm{Line Index48_{H\delta}} \le 8.0$, or
\begin{equation}
M_{G(A-type)} = -0.131\times \rm{Line Index48_{H\delta}}+3.620
\end{equation}
if $\rm{Line Index48_{H\delta}}>8.0$.
The $1\sigma$ uncertainty of the derived $M_{G}$ is taken from the residual of the fitting, which is 0.60 mag, corresponding to a precision of 30\% in distance.

By combining the $E(B-V)$ values from \citet{green2019ApJ}, the $G$-band absolute magnitudes $M_{G}$ derived from the spectra, and the {\it Gaia} DR3 $G$ magnitudes, we calculate the photometric distances and their uncertainties using the Monte Carlo method, accounting for the uncertainties in both $M_{G}$ and $G$ magnitudes. 
Once the parallax or photometric distances are determined, the 3D positions and 3D velocity components in a right-handed Cartesian system\footnote{The position of the Sun is set to ($X$, $Y$, $Z$) = ($-8.34$, $0$, $0$)~kpc.} and $v_{\rm{tot}}$ can be re-estimated. Their uncertainties are again calculated through the Monte Carlo Method by considering the Gaussian uncertainties of the distances, proper motions and radial velocities. 

Among the 86 objects without reliable \textit{rgeo} as indicated in Table~\ref{tab:selection}, photometric distances of all but 18 have been re-determined. The 18 objects either do not have a good spectral type fit by the MKCLASS code, or are not applicable to the line index method due to being non-A type or having low SNR. Using the photometric distances, we confirm that all of the eight potential HVS candidates with $v_{\rm{rf}} < 300$ km s$^{-1}$ have high total velocities larger than 440 km s$^{-1}$. 

\section{Results and Discussion} \label{sec:result}
\begin{figure*}
	\begin{centering}
		\includegraphics[width=1.98\columnwidth]{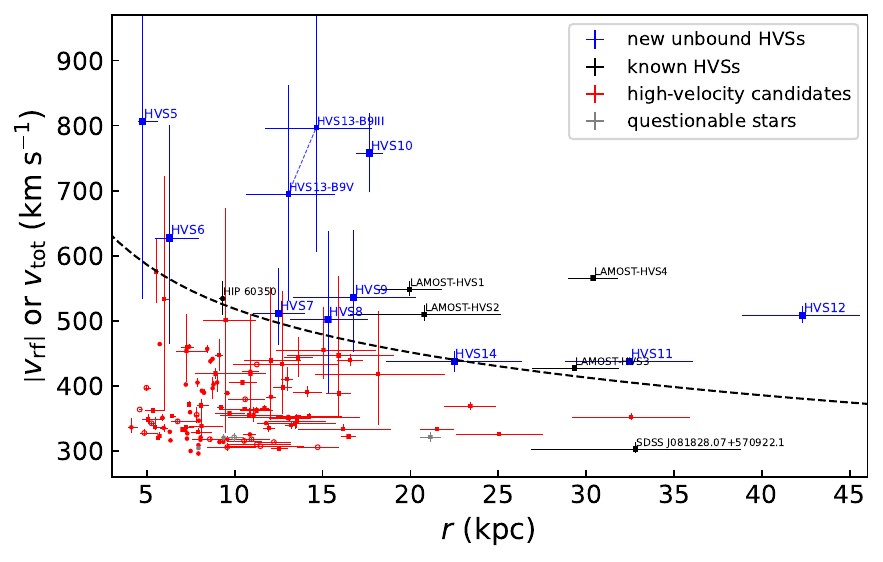}
		\caption{The velocities versus Galactocentric distances of the 125 high-velocity stars in our results are plotted. Blue, black, grey, and red symbols indicate the ten new unbound HVS candidates (from LAMOST-HVS5 to LAMOST-HVS14), six known HVSs, high-velocity star candidates with questionable velocities, and the remaining candidates, respectively. Solid circles indicate that $v_{\rm{tot}}$ derived from reliable parallax distances is used. Squares indicate that $v_{\rm{tot}}$ derived from photometric distances or from previous work is used. Empty circles indicate that $v_{\rm{rf}}$ is used due to the lack of reliable distances. The dashed curve represents the escape velocity curve using the MWPotential2014 \citep{2015ApJS..216...29B}. }
		\label{fig:v}
	\end{centering}
\end{figure*}

We present the properties of the sample of HVS candidates as well as individual objects in this section. After manually adding LAMOST-HVS2 to the search result, the final high-velocity candidate sample contains 125 stars, including 117 stars with $\left| v_{\rm{rf} } \right| > 300$ km s$^{-1}$ and eight stars with $v_{\rm{tot}} > 440$ km s$^{-1}$ but $\left| v_{\rm{rf} } \right| < 300$ km s$^{-1}$. For more precise velocity measurements, we recalculated the velocities using the radial velocities from the LAMOST AFGK-stellar parameter catalog (which can be obtained from the LAMOST website\footnote{\url{https://www.lamost.org/dr10/v1.0/catalogue}}) with a radial velocity zero-point correction of +4.574 km s$^{-1}$ \citep{2023RAA....23k5026L}. Four stars in the final sample are marked with questionable radial velocities measurement by eye check. In addition, 20 stars show significant periodic light variations in ZTF $g$/$r$ bands. The final sample containing the 125 high-velocity star candidates is presented in Table~\ref{tab:hivel}. We demonstrate the Galactocentric distances and velocities for the final sample in Figure~\ref{fig:v}. We begin by describing the properties of individual interesting stars, followed by a discussion of the overall distribution of the sample.

\begin{figure*}
	\begin{centering}
		\includegraphics[width=2.0\columnwidth]{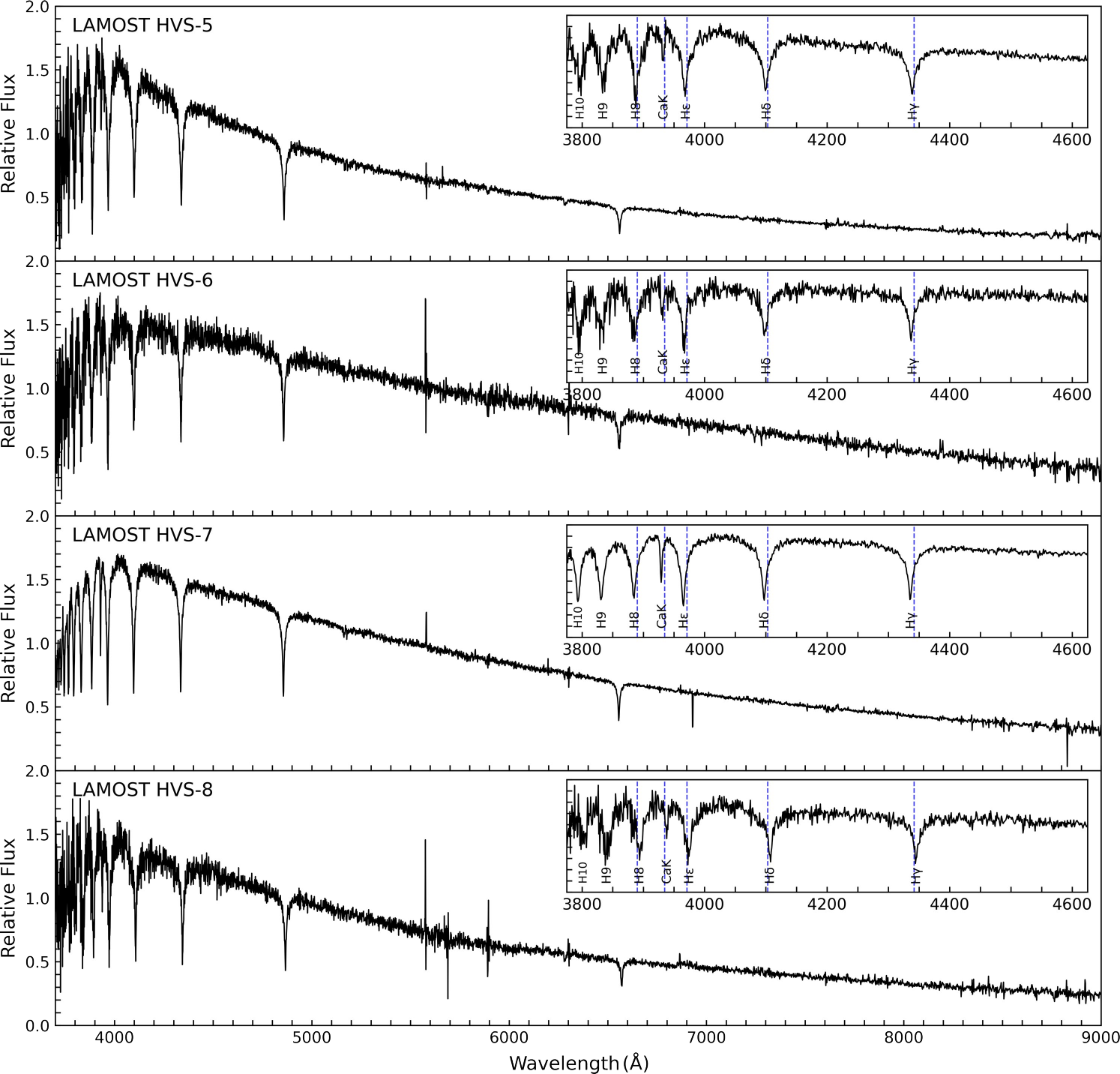}
		\caption{LAMOST spectra of the new unbound HVS candidates LAMOST-HVS5, 6, 7 and 8. The inset in each panel presents an enlarged view of the normalized blue-arm spectrum, offering a clearer view of the detected absorption lines. The vertical dashed lines mark the rest wavelengths of these lines.}
		\label{fig:hvs1}
	\end{centering}
\end{figure*}

\begin{figure*}
	\begin{centering}
		\includegraphics[width=2.0\columnwidth]{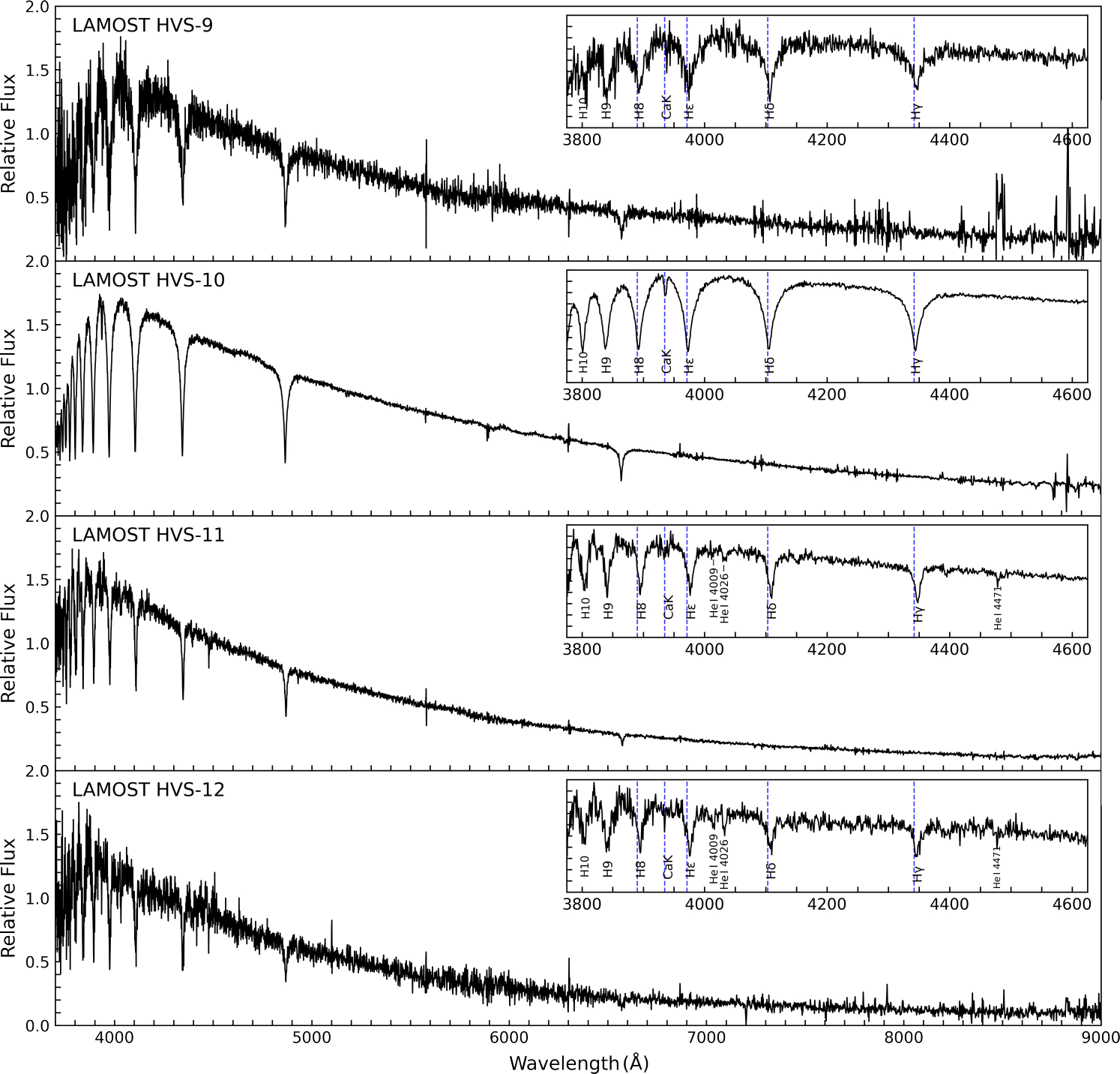}
		\caption{Similar to Figure~\ref{fig:hvs1} but for HVS candidates LAMOST-HVS9, 10, 11 and 12.}
		\label{fig:hvs2}
	\end{centering}
\end{figure*}

\begin{figure*}
	\begin{centering}
		\includegraphics[width=2.0\columnwidth]{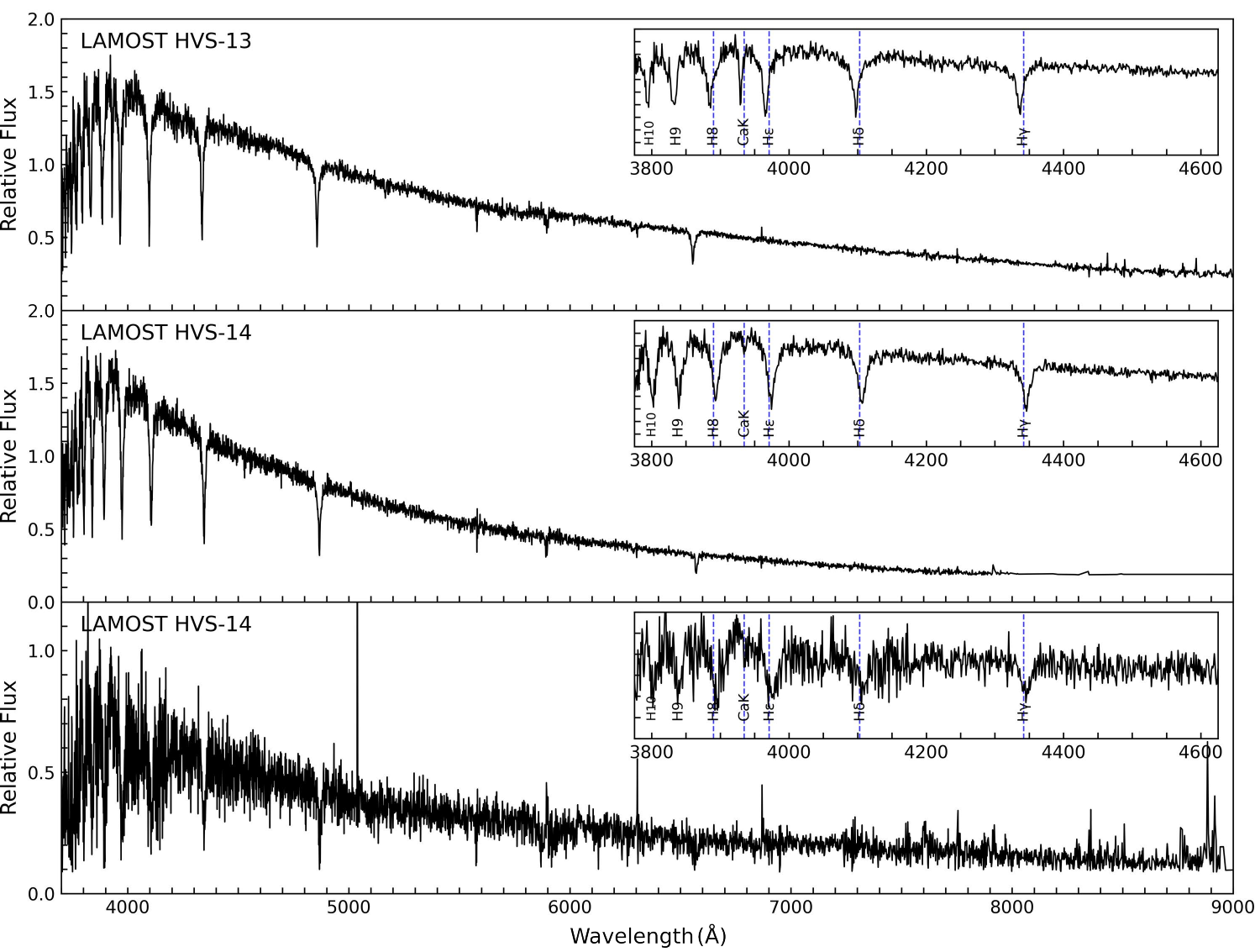}
		\caption{Similar to Figure~\ref{fig:hvs1} but for HVS candidates LAMOST-HVS13 and 14. For LAMOST-HVS14, we present the two single LAMOST observation separately (the higher-SNR spectrum in the middle pannel and the lower one in the bottom pannel).}
		\label{fig:hvs3}
	\end{centering}
\end{figure*}

\subsection{Individual interesting objects}\label{sec:individual}
\subsubsection{New unbound HVS candidates}\label{sec:newhvs}
Through a comparison between the velocities of our entire sample and the Milky Way's escape velocity curve \citep{2015ApJS..216...29B}, we identified ten new unbound HVSs.
Hereafter, they are denoted by LAMOST-HVS5 to LAMOST-HVS14. Their spectra are presented in Figures~\ref{fig:hvs1} and \ref{fig:hvs2}, with detailed parameters listed in Tables~\ref{tab:hvs} and \ref{tab:hvs9}. After examining the light curves from ZTF data, we found no significant variability for these HVS candidates except for LAMOST-HVS5 and LAMOST-HVS13, whose characteristics are discussed below.

\begin{figure*}
	\begin{centering}
		\includegraphics[width=1.9\columnwidth]{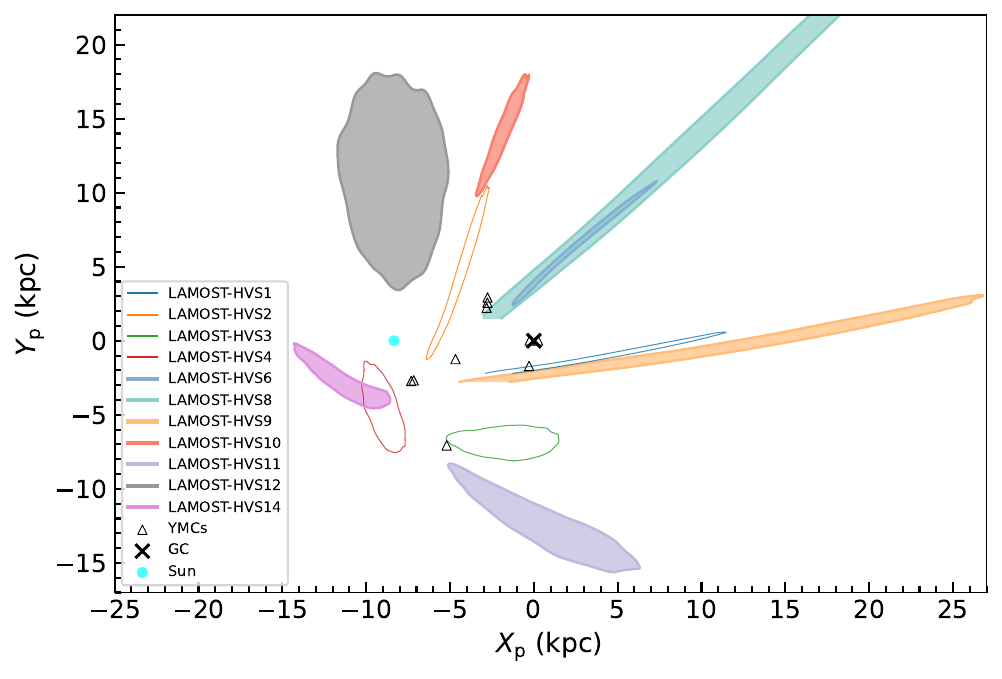}
		\caption{The contours indicate the intersection regions in Galactic $X$, $Y$ plane of the backward-integrated orbits and the Galactic disc for each star. LAMOST-HVS1 to LAMOST-HVS4 are marked by unfilled contours, while the new LAMOST HVSs are marked by filled contours. The black cross and the cyan circle indicate the position of the Galactic center and the sun, respectively. The triangles mark the known young massive clusters (YMCs) taken from \citet{2010ARA&A..48..431P}.}
		\label{fig:cross}
	\end{centering}
\end{figure*}

To investigate the origins of these HVSs, we used the \textsc{Python} package \textsc{galpy} \citep{2015ApJS..216...29B} for backward orbital integrations. We adopted a solar Galactocentric distance of $R_0 = 8.34$ kpc \citep{reid2014ApJ} and a solar circular velocity of 230 km s$^{-1}$ \citep{2023ApJ...946...73Z} as well as the solar peculiar motion ($U_\odot$,$V_\odot$,$W_\odot$) = (11.69, 10.16, 7.67) km~s$^{-1}$ \citep{wang2021MNRAS}. These parameters are consistent with those used to calculate stellar velocities with \textsc{astropy} as described in Section~\ref{sec:target}. In our orbital integration, we employ the MWPotential2014 model, which is incorporated within the \textsc{galpy} package. 
This potential model realistically characterizes the Galaxy's mass distribution and is recommended by \textsc{galpy}. We calculated the intersection points of each Monte Carlo sampled orbit with the Galactic disk. Then, we determined the region encompassing $90 \%$ of the probability density as the disk intersection region. The final results are shown in Figure~\ref{fig:cross}.

\textit{LAMOST-HVS5.} %821307118
This star, designated as J161649.39-030624.9 in the LAMOST database, has a single observation, as shown in Figure~\ref{fig:hvs1}. Its spectrum is consistent with a A0III star as given by MKCLASS with a heliocentric radial velocity of $-207.48 \pm 6.99$ km s$^{-1}$. 
From the H$\delta$ line index, the absolute magnitude is estimated to be $2.01 \pm 0.60$, yielding a photometric distance of $7.85 \pm 2.20$ kpc. 
Located at $4.77 \pm 0.58$ kpc from the GC, it lies $Z = 4.18 \pm 1.16$ kpc above the Galactic plane. With a total velocity of $806.1 \pm 276.6$ km s$^{-1}$, significantly exceeding the local escape velocity, as depicted in Figure~\ref{fig:v}. The rotation velocity component $V = -766.6 \pm 276.6$ km s$^{-1}$ indicates that the star is moving extremely fast in the opposite direction of the Galactic rotation. It has a vertical velocity of $W = -233.1 \pm 36.6$ km s$^{-1}$ toward the south Galactic pole, indicating that the star is moving from the north toward the disk plane. 
Its radial velocity component is toward the GC. The high total velocity in this inward direction makes it unlikely that this star originated from the GC.
It is also less likely that it is a runaway star from a core-collapse supernova explosion, as the ejected energy typically accelerates stars to a maximum of 300–400 km s$^{-1}$ \citep{2000ApJ...544..437P}.
It is also hard to be ejected from the classical dwarf galaxies, such as Sgr dSph \citep{huang2021ApJ, 2023Li_AJ}, given its massive and young nature.
We also rule out an LMC origin, as its backward orbit is hardly linked to that of the LMC. Therefore, the origin of LAMOST-HVS5 remains unclear.
One possible explanation is that LAMOST-HVS5 was ejected from a young massive cluster (YMC) after experiencing strong gravitational interactions involving three to four stellar objects. 
The upper limit of the ejection velocity from these encounters is approximately equal to the escape speed from the surface of the most massive star involved \citep{1991AJ....101..562L}. For a solar-like star, this speed is about 620 km s$^{-1}$, while it can reach up to 1400 km s$^{-1}$ for a massive star with a mass of 60 M$_\odot$ \citep{1991AJ....101..562L}.
Given the young and massive nature of LAMOST-HVS5, it is possible that its companion is also a massive star. In this case, such a high velocity could be naturally explained by this dynamical interaction scenario.
Interestingly, we observe a periodic variation of 1.1 hours for LAMOST-HVS5 in the ZTF $g$- and $r$-band photometry, with a variation amplitude of 2–3\%.
Given its A-type classification and absolute magnitude, it is likely a $\delta$ Scuti star \citep{1979PASP...91....5B}. Since the typical radial velocity variation for $\delta$ Scuti stars ranges from 5 to 10 km s$^{-1}$ \citep{1976ApJ...210..163B,2018MNRAS.475..478Q}, the radial pulsation is expected to have only a minimal effect on the identification of its hypervelocity nature.

\textit{LAMOST-HVS6.} %664314064
This star, designated as J171230.82+110924.5 in the LAMOST database, has a single observation, as shown in Figure~\ref{fig:hvs1}. The best-fit spectral type is A9V, but we consider it to be questionable because the strength of H$\delta$ is less and the widths of H$\gamma$ and H$\delta$ seem to be slightly narrower than the fitted A9V template. The relatively low SNR of the LAMOST spectrum may contribute to these discrepancies. Given the intensity of the Ca\,II K line being less than that of H$\epsilon$, we infer that the spectral type is likely earlier than A9. Consequently, we rely on the H$\delta$ line index to estimate the absolute magnitude of $2.44 \pm 0.60$, corresponding to a photometric distance of $9.54 \pm 2.68$ kpc. At this distance, its total velocity ($v_{\rm{tot}} = 627.4 \pm 168.5$ km s$^{-1}$) exceeds the local escape velocity at $r = 6.29$ kpc, qualifying it as an unbound HVS candidate. The star exhibits a velocity component directed toward the northern Galactic pole and is positioned $4.29 \pm 1.20$ kpc above the Galactic plane. This suggests that it has previously crossed the Galactic plane, as illustrated in Figure~\ref{fig:cross}. For LAMOST-HVS6, the backward-integrated orbit indicates that the disk intersection region is not distant from the GC, suggesting a possible GC origin. 
The dynamical interaction scenario is also possible as some known YMCs are close to the disk intersection region, so this star may be ejected through few-body interactions from the dense region of the YMCs with a massive object involved. Considering the significant vertical velocity at the earlier time, the ejection by a core-collapse supernova becomes unlikely. Similar to LAMOST-HVS5, origins within the LMC or Sgr dSph are improbable. Thus, we conclude that this star may be ejected from either the GC or dynamical interactions.

\textit{LAMOST-HVS7.} %620008146, 643910098(not in the sample)
LAMOST-HVS7 is designated as J110108.23+360918.0 in the LAMOST database, which has three observations. However, one spectrum exhibits abnormal data due to bad data processing, so we discard it. The averaged spectrum from the remaining two observations is shown in Figure~\ref{fig:hvs1}. Given its spectral type of A0III by MKCLASS, we estimated the absolute magnitude using the H$\delta$ line index from the spectrum with the highest SNR to minimize measurement errors. This yielded an absolute magnitude of $2.30 \pm 0.60$ and a corresponding photometric distance of $6.61 \pm 1.85$ kpc. The radial velocities of the two good-quality spectra are consistent with each other, which are $-412.51 \pm 5.24$ km s$^{-1}$ and $-421.75 \pm 12.23$ km s$^{-1}$, respectively. We adopt the mean radial velocity value of $-417.13 \pm 6.65$ for determining kinetic parameters.
The total velocity of LAMOST-HVS7 includes a significant vertical component $W = -498.8 \pm 36.5$ km s$^{-1}$ directed toward the southern Galactic pole. With a positive vertical distance of $Z = 6.06 \pm 1.66$ kpc above the disk plane, LAMOST-HVS7 exhibits an inward orbit similar to that of LAMOST-HVS5. This orbit rules out the GC origin. Given such a high vertical velocity component, the core-collapse supernova ejection scenario becomes unlikely. Origins within the LMC or Sgr dSph can also be ruled out for the same reasons as in LAMOST-HVS5. Thus, LAMOST-HVS7, like LAMOST-HVS5, is suggested to originate from the dynamical interaction scenario.

\textit{LAMOST-HVS8.} %868007134
This star, designated as J023346.69+103930.7 in the LAMOST database, has a single observation, as shown in Figure~\ref{fig:hvs1}. Its spectrum fits well with an A0III template from MKCLASS. Using the H$\delta$ line index, we estimate the absolute magnitude to be $2.38 \pm 0.60$, yielding a photometric distance of $8.68 \pm 2.43$ kpc. The total velocity $v_{\rm{tot}} = 501.6 \pm 124.2$ km s$^{-1}$ exceeds the local escape velocity at a Galactocentric distance of $r = 15.30 \pm 2.21$ kpc. LAMOST-HVS8 is currently moving away from the disk plane and its orbit likely intersected with the Galactic disk in the past. From Figure~\ref{fig:cross}, the backward-integrated orbit of LAMOST-HVS8 intersects the disk near the GC direction and close to the known YMCs, specifically RSGC1-3 \citep{2010ARA&A..48..431P}. Given its southward velocity direction, origins within the LMC or Sgr dSph can be ruled out. The relatively high ejection velocity also makes the core-collapse supernova ejection scenario less plausible. 

\textit{LAMOST-HVS9.} %298811067
This star, designated as J125538.89+352338.8, has a single observation in the LAMOST database, as shown in Figure~\ref{fig:hvs2}. The spectrum is consistent with an A1IV star as given by MKCLASS. Using the H$\delta$ line index, we estimate the absolute magnitude to be $1.68 \pm 0.60$, yielding a photometric distance of $14.16 \pm 3.94$ kpc. LAMOST-HVS9 is located at a high Galactic latitude ($b=81.7^{\circ}$), indicating a large vertical distance of $13.99 \pm 3.98$ kpc above the disk plane. Its total velocity is estimated to be $535.6 \pm 93.6$ km s$^{-1}$, which is larger than the local escape velocity. The velocity direction indicates that this star is moving away from the Galactic disk plane. Given its significant vertical distance north of the disk plane, it is unlikely that this star is associated with either the LMC or Sgr dSph. From Figure~\ref{fig:cross}, the disk intersection region of the backward-integrated orbit of LAMOST-HVS9 is near the GC and a known YMC, i.e., (DBS2003) 179 \citep{2010ARA&A..48..431P}. The substantial vertical velocity component makes the supernova ejection scenario less plausible.

\begin{deluxetable*}{cccccccccccc}
\tabletypesize{\scriptsize}
\tablecaption{125 high-velocity candidates from LAMOST DR10\label{tab:hivel}}
\tablehead{
\colhead{Designation} & \colhead{Name} & \colhead{MKCLASS} 
& \colhead{RA} & \colhead{DEC} & \colhead{Gmag}  
& \colhead{Vrf} & \colhead{Vtot} & \colhead{r} 
& \colhead{dist\_source} & \colhead{RV\_questionable} & \colhead{Var}\\
\nocolhead{} & \nocolhead{} & \nocolhead{} 
& \colhead{(deg)} & \colhead{(deg)} & \colhead{(mag)}  
& \colhead{(km s$^{-1}$)} & \colhead{(km s$^{-1}$)} & \colhead{(kpc)} 
& \nocolhead{} & \nocolhead{} & \nocolhead{}
} 
%\colnumbers
\startdata 
    J000501.67+174815.0 & ~ & A9V? & 1.25694 & $17.80414$ & 17.63 & 305.5$^{+1.2}_{-1.2}$ & 417.9$^{+97.8}_{-77.4}$ & 18.17$^{+3.79}_{-3.60}$ & A\_index & ~ & ~ \\
    J001112.03-041947.5 & ~ & A4III & 2.80017 & $-4.32990$ & 17.28& 399.9$^{+5.8}_{-5.9}$& 454.7$^{+66.0}_{-43.9}$ & 15.05$^{+2.96}_{-2.71}$ & A\_index  & ~ & Y \\
    J005837.09+445430.8 & ~ & ~ & 14.65455 & $44.90862$ & 16.79 & 378.0$^{+18.3}_{-18.4}$ & 396.4$^{+17.1}_{-17.1}$ & 8.74$^{+0.04}_{-0.04}$ & rgeo & ~ & Y \\
    ... & ... & ... & ... & ... & ... & ... & ... & ... & ... & ... & ...\\
    J234029.60-002300.8 & ~ & ~ & 355.12319 & $-0.38360$ & 13.34 & 328.3$^{+2.1}_{-2.1}$ & 456.9$^{+6.2}_{-6.0}$ & 8.44$^{+0.04}_{-0.04}$ & rgeo & ~ & ~ \\
    J234921.86+090714.5 & ~ & ~ & 357.34117 & $9.12072$ & 17.61 & 354.8$^{+3.0}_{-2.9}$  & 392.6$^{+2.6}_{-2.6}$ & 8.13$^{+0.00}_{-0.00}$ & rgeo & ~ & ~ \\
\enddata
\tablecomments{
Column (1) - Designation of the LAMOST target.
Column (2) - Star name for LAMOST HVSs and other known HVSs.
Column (3) - Spectral classification given by MKCLASS code.
Column (4) - GAIA DR3 right ascension.
Column (5) - GAIA DR3 declination.
Column (6) - GAIA DR3 $G$-band magnitude.
Column (7) - Radial velocity in the Galactic rest-frame.
Column (8) - Total velocity in the Galactic rest-frame.
Column (9) - Galactocentric distances.
Column (10) - Sources from which the stellar distances are obtained. `rgeo' for Gaia parallax distances, `A\_index' for photometric distances derived from line indexes, `MK' for photometric distances derived from the stellar spectral classification, `Brown07' for distances taken from \citet{brown2007hyper}, `Huang17' for distances taken from \citet{huang2017}, `Li18' for distances taken from \citet{li2018}, and `He25' for distances taken from \citet{2025He}.
Column (11) - `Y' for stars with questionable radial velocity measurements.
Column (12) - `Y' for stars showing obvious light variation.
The complete version of this table is available online.
}
\end{deluxetable*}

\textit{LAMOST-HVS10.} %505106037 497306037 (629601070)
LAMOST-HVS10 is designated as J015725.24+323539.3 in the LAMOST database, which has three observations. The best fit of the spectra is A0V as given by MKCLASS. The radial velocities from the three observations are $169.48 \pm 7.12$ km s$^{-1}$, $183.46 \pm 5.84$ km s$^{-1}$, and $156.34 \pm 5.12$ km s$^{-1}$, respectively. Given the lack of significant light variations, these discrepancies are more likely attributable to measurement uncertainties rather than stellar pulsations or binary motion. The photometric distance, estimated from the absolute magnitude of an A0V star, is $11.20 \pm 0.82$ kpc. Using the averaged $v_{\rm{los}}$ of $169.8 \pm 2.9$ km s$^{-1}$ and the photometric distance, we derive a total velocity of $758.0 \pm 61.3$ km s$^{-1}$, far exceeding the local escape velocity at its Galactocentric distance of $r = 17.67 \pm 0.76$ kpc. Located $5.27 \pm 0.38$ kpc south of the Galactic plane, LAMOST-HVS10 is moving toward the Galactic south pole at a speed of $W=-235.5 \pm 12.1$ km s$^{-1}$. 
Therefore, LAMOST-HVS10 is moving away from the GC and the disk with a large speed. As illustrated in Figure~\ref{fig:cross}, its backward-integrated orbit intersects the disk near the GC, similar to LAMOST-HVS2, so the origin of the GC or the dynamical interaction cannot be excluded. The total velocity is too high for a core-collapse supernova origin. The young and massive nature of this star disfavors a possible Sgr dSph origin. The LMC origin is also unlikely considering its velocity direction.

\textit{LAMOST-HVS11.} %43510220 551911175 34116182(not in the sample)
LAMOST-HVS11 is designated as J101018.82+302028.1 in the LAMOST database, which has three observations. These observations all showing consistent line-of-sight radial velocities from 410 to 430 km s$^{-1}$. The averaged spectrum of the three observations is shown in Figure~\ref{fig:hvs2}. The proper motion of this star is quite small, so the total velocity is dominated by the radial velocity. Classified as a B5V star by MKCLASS, the photometric distance is estimated to be $27.29 \pm 3.75$ kpc, yielding a total velocity of $437.0 \pm 4.2$ km s$^{-1}$, which exceeds the local escape velocity. LAMOST-HVS11 is located at a large vertical distance of $Z=22.39 \pm 3.07$ kpc north of the disk plane and moves toward the north Galactic pole at a speed of $W = 279.6 \pm 10.2$ km s$^{-1}$. This excludes the possible relation with LMC and Sgr dSph as the star needs to travel through the disk and its flight time will be larger than the main-sequence lifetime. Although the total velocity is not as high as the other HVS candidates, the core-collapse supernova ejection scenario is hard to explain the high $Z$. From Fig. \ref{fig:cross}, although the backward orbit suffers from uncertainties in distance determination and astrometry solutions, the intersect positions still shows potential relations with the GC and known YMC, indicating the origins from GC or through dynamical interactions.

\textit{LAMOST-HVS12.}  %779510173
This star, designated as J014106.87+163157.3, has a single observation in the LAMOST database, as shown in Figure~\ref{fig:hvs2}. Given its small proper motion, the total velocity is dominated by the radial velocity $v_{\rm{rf}} = 488.3 \pm 3.9$ km s$^{-1}$ directing away from the GC. (This $v_{\rm{rf}}$ is the second largest in our sample (the largest belongs to LAMOST-HVS4).) Its spectrum is well fitted by a B3V template as given by MKCLASS. The theoretical absolute magnitude for B3V stars ($M_{G} = -1.19 \pm 0.20$) yields a heliocentric distance of $37.28 \pm 3.42$ kpc. Located at a vertical distance of $-26.20 \pm 2.39$ kpc south of the disk plane, LAMOST-HVS12 is moving toward the south Galactic pole. As illustrated in Figure~\ref{fig:cross}, although the disk intersection region is quite large, it does not significantly deviate from the direction of the GC. Thus, the relation with the GC and YMCs cannot be excluded, given the sensitivity of orbit integration to the uncertainties in distance and astrometry at such a large Galactocentric distance. Supernova ejection from the disk cannot impart such high momentum to eject this star to this large distance with its current high speed. The young and massive nature of LAMOST-HVS12 makes an Sgr dSph origin impossible, and its southward velocity makes it unlikely to be associated with the LMC. 

\textit{LAMOST-HVS13.}
%593108017
The star J212520.96+190802.1, hereafter referred to as LAMOST-HVS13, has only one observation in the LAMOST database. The spectrum is best fitted by a B9III template. Assuming the absolute magnitude for a B9V star ($M_{G} = 0.515 \pm 0.525$), the calculated total velocity is $690 \pm 162$ km s$^{-1}$. Adjusting this value to match the best-fitted B9III star HD175640, with an absolute magnitude of $M_{G} = 0.205 \pm 0.525$ (we simply continue to use the uncertainty of the main-sequence absolute magnitude), increases the total velocity to $797 \pm 192$ km s$^{-1}$ and the Galactocentric distance from 13.0 to 14.6 kpc. In this case, the velocity components will be $U = 672 \pm 190$ km s$^{-1}$, $V = -361 \pm 61$ km s$^{-1}$, and $W = 231 \pm 19$ km s$^{-1}$. In both solutions, the total velocity exceeds the local escape velocity, as illustrated in Figure~ \ref{fig:v}. Given its vertical distance of $Z = -5.60 \pm 1.35$ kpc (Figure~\ref{fig:xyz}), the orbit of LAMOST-HVS13 is directed inward, making a GC origin unlikely. Similarly to LAMOST-HVS5 and 7, we exclude a core-collapse supernova or LMC origin, while we cannot exclude a dynamical ejection scenario. A better determination of the photometric distance is required to establish a more precise backward orbit integration. Furthermore, ZTF data reveal that LAMOST-HVS13 exhibits a periodic light variation with a period of 1.55 hours. This star has been classified as a $\delta$ Scuti variable by \citet{2020ApJS..249...18C}. If this star is indeed a $\delta$ Scuti star, the small-amplitude radial velocity variation of a $\delta$ Scuti-type pulsation would not affect the velocity measurement.

%368311026 405409240
\textit{LAMOST-HVS14.}
LAMOST-HVS14 is designated as J015022.75+483631.1 in the LAMOST database, which has two observations, as shown in Figure~\ref{fig:hvs3}. The higher-SNR spectrum is well fitted by a B8V template according to MKCLASS, while the second observation, with very low SNR, does not provide a reliable fit. Based on the absolute magnitude for B8V stars of $M_{G} = -0.01 \pm 0.53$, we estimate a photometric distance of $16.3 \pm 4.0$ kpc. As the proper motion of this star is small, the total velocity is dominated by the radial alone. The two LAMOST spectra exhibit a velocity difference of approximately 60 km s$^{-1}$. Specifically, the higher-SNR spectrum yields a velocity of $254.0 \pm 0.7$ km s$^{-1}$, whereas the lower-SNR spectrum suggests a significantly higher velocity of $315.3 \pm 2.1$ km s$^{-1}$. This discrepancy is confirmed through cross-correlation analysis across the entire spectral range. As the H$\alpha$ region is more affected by noise, we applied Gaussian fitting to the H$\beta$ and H$\gamma$ lines of each spectrum. We derived average radial velocities of $254.7 \pm 16.6$ km s$^{-1}$ and $299.7 \pm 45.2$ km s$^{-1}$ for the higher- and lower-SNR spectra, respectively. Given these radial velocities, the calculated total velocities are $438.0 \pm 16.1$ km s$^{-1}$ and $480.6 \pm 43.9$ km s$^{-1}$. At a Galactocentric distance of $r = 22.5 \pm 3.9$ kpc, where the escape velocity is around 440 km s$^{-1}$, the radial velocities implies that the star might be unbound from the Galaxy, as indicated in Figure~\ref{fig:v} with $v_{\rm{tot}}$ = $438.0 \pm 16.1$ km s$^{-1}$. We consider the parameters derived from the velocity of the higher-SNR spectrum to be more reliable. We present the parameters derived from the velocities from the spectral line fitting of the higher-SNR spectrum in Table~\ref{tab:hvs13}. Based on these parameters, backward orbit integration indicates that the disk intersection region is opposite to the direction of the GC, making a GC origin less likely. We consider that LAMOST-HVS14 could be ejected from the core-collapse supernova or dynamical ejection scenario. However, because of the limited data of only two epochs, we cannot definitively conclude whether the observed velocity changes are genuine. Such variations could potentially arise from binary motion, but we did not detect significant variability from ZTF light curves. Furthermore, the {\it Gaia} DR3 \textit{RUWE} is 1.05 for this star, which does not suggest the presence of a wide binary companion. Thus, we currently interpret the velocity difference as possibly being a measurement error. Nevertheless, even if it is proven to be a binary star system, it may possess a high systematic velocity, which is also rare and interesting. Further spectroscopic observations are necessary to confirm whether the velocity variations are due to stellar pulsation and whether the star is a true HVS.

\subsubsection{Special stars}\label{sec:sp}
The only WD in our sample J234921.86+090714.5 has only one spectrum in the LAMOST database, showing DA-type features. It has a heliocentric distance of $86.8  \pm 1.1$ pc. Its location in the {\it Gaia} CMD (Figure~\ref{fig:cmd}) is below the main white dwarf cooling track and the broad Balmer lines indicate that this star is a massive WD. No significant light variation is detected in the ZTF data, which indicates that this is likely a single WD. According to the catalog of nearby high-mass white dwarfs from \citet{2019ApJ...886..100C}, the mass of this star is $1.18 M_{\odot}$ assuming a C/O core or 1.14 $M_{\odot}$ assuming an O/Ne core, and the age is estimated to be 1.12 Gyr in both cases. This star is $46.4 \pm 0.8$ pc south from the Galactic plane with a vertical velocity $W = -155.05 \pm 2.29$ km s$^{-1}$ toward the south Galactic pole. The Galactocentric radial velocity component $U = -17.69 \pm 0.30$ km s$^{-1}$ is small, and the velocity component in the direction of rotation is $V = 130.21 \pm 1.85$ km s$^{-1}$. Therefore, the star's orbit is not planar. \cite{Raddi2022AA} investigated the kinematic properties of about 3300 white dwarfs and identified the WD LSPM J1756+0931N with a large total velocity of $316.5$ km s$^{-1}$ as a halo candidate. We propose that J234921.86+090714.5 may be also a relatively extreme halo WD member similar to LSPM J1756+0931N. It is also possible that it comes from the multi-star interaction or a stellar merger. Further orbital analysis is required to confirm whether it is a halo member or it has other origins. 

J153747.37+034101.2 is a type-c RR Lyrae star with $v_{\rm{rf}} = 436.3 \pm 2.3$ km s$^{-1}$ found in our sample, which is presented in the RR Lyrae catalog from \citet{2013AJ....146...21S}. We find its period of 6.731\,h in the ZTF data. The two spectra from LAMOST DR10 both show A-type star features with line-of-sight radial velocities of more than 390 km s$^{-1}$. 
Although its $rgeo$ distance is $11.0$ kpc which is considered to be unreliable, as an RR Lyrae star, we can determine its distance from its pulsation properties. The distance can be directly obtained through the application of the period-Wesenheit-metallicity (PWZ) relation for RR Lyrae stars based on three-band ZTF data. Through the procedure described in \citet{2025He}, the heliocentric distance of this RR Lyrae star is estimated to be $9.98 \pm 0.82$ kpc, consistent with the distance of 9.99 kpc from \citet{2013AJ....146...21S}.
Based on our distance, the star's total velocity is estimated to be $459.0 \pm 3.7$ km s$^{-1}$. As the radial velocity amplitude for RRc stars is typically several tens of km s$^{-1}$ \citep{2021ApJ...919...85B}, the system velocities will not be affected within several tens of km s$^{-1}$. This star is located at north of the GC, with a 6.9 kpc vertical distance to the Galactic plane. The three-dimensional velocity components are $U=408.4 \pm 9.9$, $V=71.9 \pm 17.8$ and $W=196.3 \pm 7.1$ km s$^{-1}$, so this star is now moving away from the disk plane.

\subsubsection{Stars with questionable velocities}\label{sec:bound}

We mark four stars in our final sample with questionable velocity measurements. They are influenced by obvious radial velocity variations amongst multiple observations or low SNR in their spectra. The spectra of J130351.83+640709.8 have very low SNRs in the blue part, so we consider the radial velocity to be questionable.
The two spectra of J073021.28+350045.2 in the LAMOST database give an obvious deviation in the radial velocity, which may be affected by a possible binary nature or a systematic error. 
J112227.84+244341.3 also have a relatively low SNR and have a difference in the radial velocity between two observations.
The spectrum of J205520.46+334443.5 has a very low SNR so the radial velocity is not reliable.

\subsection{Distributions of the high-velocity candidates}\label{sec:distr}
\begin{figure*}
	\begin{centering}
		\includegraphics[width=1.98\columnwidth]{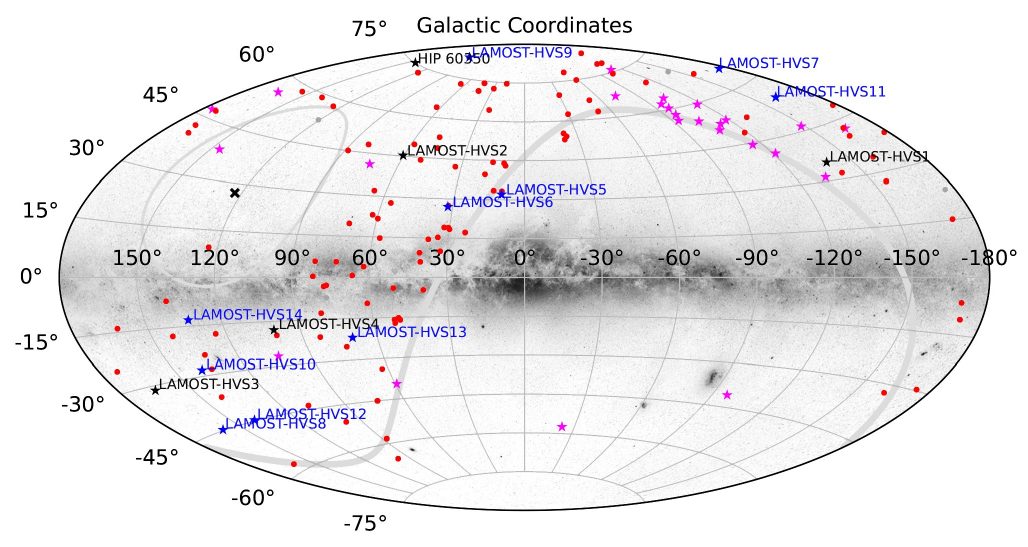}
		\caption{The spatial distribution of the LAMOST HVS candidates and previously known HVSs is shown in  Galactic coordinates $l$ and $b$ with Aitoff projection. Blue and black star symbols represent the ten new unbound HVSs and previously known HVSs found in our results, respectively. Grey and red circles indicate the HVS candidates with questionable velocities, and the remaining candidates, respectively. The 24 HVSs discovered by \citet{brown2014mmt} and S5-HVS1 \citep{2020MNRAS.491.2465K} are marked as magenta star symbles. The grey broad band marks the celestial equator and the narrow band marks declination $\delta = 60^{\circ}$. The cross marker indicates the north celestial pole.}
		\label{fig:gc}
	\end{centering}
\end{figure*}

In this section, we explore the spatial and kinematic distributions of the 125 high-velocity star candidates identified in this work.
Figure~\ref{fig:gc} illustrates the spatial distribution of all high-velocity candidates in the Galactic coordinate system as seen from the Sun. These candidates are distributed across the main survey area of LAMOST, which spans declinations from $\delta = 0 ^{\circ}$ to $60 ^{\circ}$. 
Notably, previously identified HVSs have Galactic latitude $|b| > 15^{\circ}$, indicating they are not confined within the Galactic disk. 
Similarly, the newly discovered unbound HVS candidates (LAMOST-HVS5 to LAMOST-HVS14) also exhibit large Galactic latitudes. Specifically, LAMOST-HVS5, HVS6, and HVS14 have relatively small vertical distances ($|Z| < 5$ kpc), reflected by their lower Galactic latitudes ($|b| < 35^{\circ}$). Among these, LAMOST-HVS14 has the smallest vertical distance and the lowest Galactic latitude ($|b| = 13.12^{\circ}$), though it is still above the galactic disk.

\begin{figure}
	\begin{centering}
	\includegraphics[width=1.0\columnwidth]{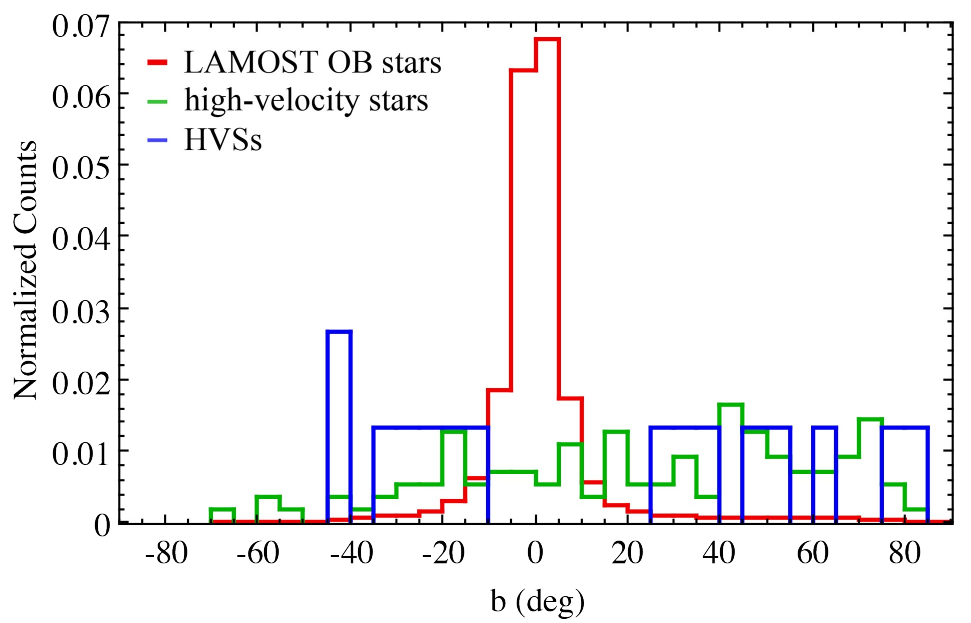}
		\caption{The distribution of the Galactic latitude $b$ for 16053 LAMOST OB stars (red), 15 unbound known HVSs/HVS candidates (blue, including LAMOST-HVS1 to LAMOST-HVS14 and HIP60350), and the other high-velocity candidates from this work (green).}
		\label{fig:b}
	\end{centering}
\end{figure}

To further examine these distributions, Figure~\ref{fig:b} compares the Galactic latitude distributions among the unbound HVSs/HVS candidates, other high-velocity candidates discovered in this work, and LAMOST OB stars \citep{2019ApJS..241...32L}. The OB star sample shows a clear peak at $|b| < 15^{\circ}$, indicative of their concentration near the galactic disk. In contrast, the HVSs/HVS candidates are more evenly distributed. Specifically, the unbound HVSs show a weak dip for $|b| < 10^{\circ}$, highlighting that their high velocities prevent them from being confined within the galactic disk.

\begin{figure*}
	\begin{centering}
		\includegraphics[width=1.75\columnwidth]{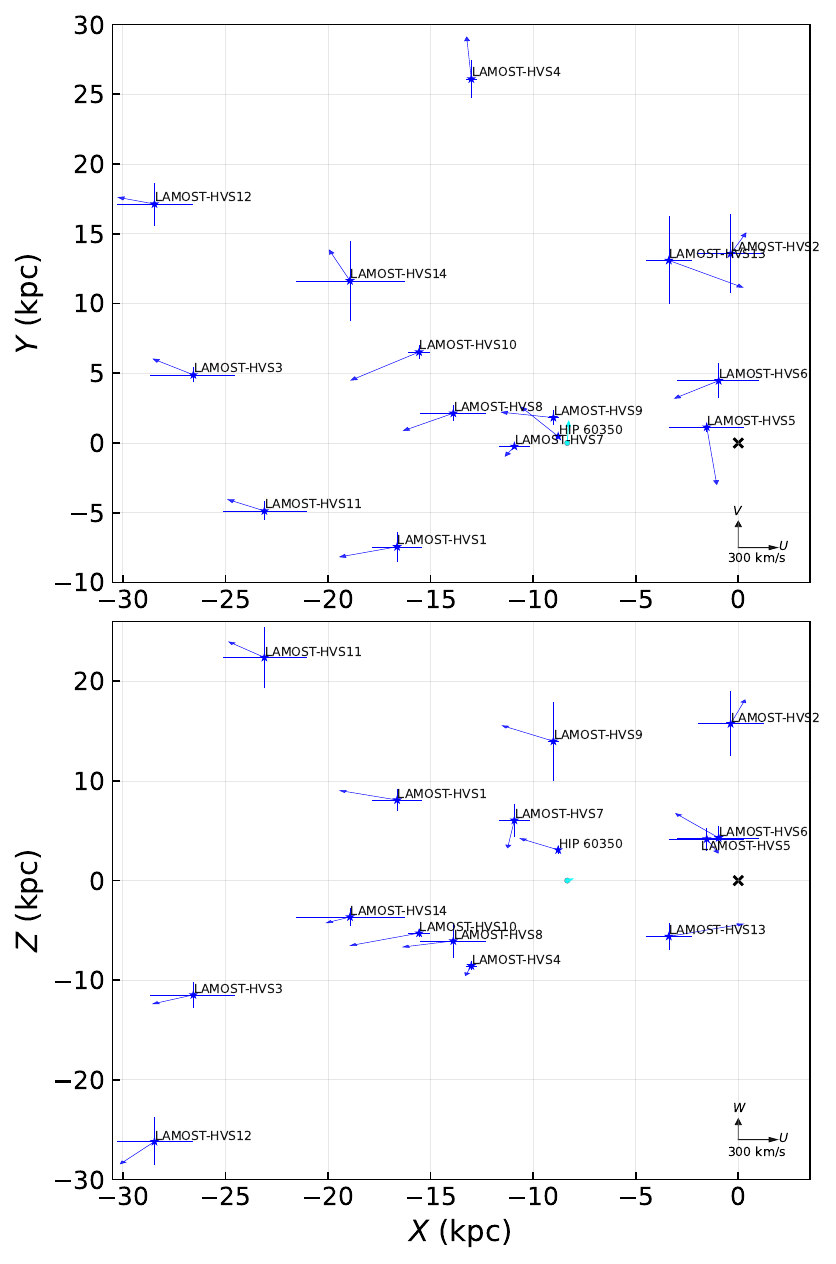}
		\caption{The locations for the HVS candidates with reliable parallax or photometric distances in our results in the Galactocentric right-handed Cartesian coordinate are shown. The Cartesian $U$, $V$, and $W$ velocity components are indicated by arrows and their values are indicated by the length of the arrows. The arrows at the right lower part of each panel indicate the length of 300 km s$^{-1}$ in both directions. The cross marker represents the location of the GC. The cyan circle and arrow mark the solar location and peculiar velocity.}
		\label{fig:xyz}
	\end{centering}
\end{figure*}

We explore the kinematic properties of these candidates, which offer critical clues about their origins — either ejected from the GC via the Hills mechanism or through dynamical interactions within YMCs. For previously known LAMOST HVSs, LAMOST-HVS1 and HVS2 face flight time issues when traced back to the GC \citep{huang2017}, suggesting a possible origin from the Galactic disk. Further analysis by \citet{2019ApJ...873..116H} confirms that LAMOST-HVS1 was ejected from the inner galactic disk through multi-body dynamical interactions. 
Conversely, LAMOST-HVS3 shows no such issue, and LAMOST-HVS4's orbit suggests a disk origin \citep{li2018}. 
From Figure~\ref{fig:cross} (and also Figure~5 in \citet{2020MNRAS.491.2465K}), the current data do not strongly support a GC origin for LAMOST-HVS3 and LAMOST-HVS4.
Figure~\ref{fig:xyz} plots the locations and velocity components in Cartesian coordinates for LAMOST-HVS1 to LAMOST-HVS14 and the unbound star HIP60350. It reveals that most HVSs have velocities directed away from the GC, except for LAMOST-HVS5, HVS7, and HVS13. The lower panel highlights the vertical location distribution, showing that most HVS candidates have the same signs between the vertical distance $Z$ and the vertical velocity component $W$. This outward pattern suggests potential origins either near the GC related to the SMBH or within the galactic disk linked to YMCs. 
Among the unbound LAMOST HVS candidates, LAMOST-HVS5, HVS7, and HVS13 exhibit inward orbits, pointing to possible birthplaces at significant vertical distances from the galactic plane.

The spectral types of the previous identified LAMOST HVSs (LAMOST-HVS1 to LAMOST-HVS4) are all B-type stars, consistent with other HVS surveys. In contrast, the new unbound HVS candidates discovered through this systematic search from LAMOST are mostly A-type stars, with four B-type stars among them. 
Our sample also includes one bound WD with a hydrogen-dominated atmosphere and a total velocity below 600 km s$^{-1}$, suggesting it might be a member of the galactic halo but not related to thermonuclear supernova explosions. 
Continuous searches for hyper-velocity WDs in future LAMOST data releases, focusing on WDs that do not have hydrogen-dominated atmospheres, are expected to uncover new hyper-velocity WDs or subdwarfs like GD 492, US 708, and the D6 stars. This will also help to set constraints on the formation channels of the hyper-velocity WD and subdwarfs that originate from thermonuclear supernova explosions.

\begin{table*}
	%\footnotesize
	\scriptsize
	\centering
	\caption{Properties of the new unbound HVS candidates LAMOST-HVS5-8}
	\label{tab:hvs}
 \makebox[\textwidth][c]{
    \begin{threeparttable}
	\begin{tabular}{lcccc} % four columns, alignment for each
		\hline
		\hline
		& LAMOST-HVS5 & LAMOST-HVS6 & LAMOST-HVS7 & LAMOST-HVS8 \\
		\hline
		LAMOST designation & J161649.39-030624.9 & J171230.82+110924.5 & J110108.23+360918.0 & J023346.69+103930.7\\
		Position ($\alpha^{\circ}, \delta^{\circ}$) (ICRS(Epoch J2016.0)) 
		& (244.205784,-3.106973) 
		& (258.128288,11.156784) 
		& (165.284255,36.154960) 
		& (38.444620,10.658501)\\
		($l^{\circ}, b^{\circ}$) 
		& (9.830464,32.030157) 
		& (31.970658,26.970319) 
		& (184.933861,64.956339) 
		& (159.660458,-44.791921) \\
		Magnitudes \tnote{a} 
		 & $G=16.909\pm0.003$ & $G=17.687\pm0.003$ & $G=16.406\pm0.003$ & $G=17.300\pm0.003$\\
		 & $g=17.060\pm0.005$ & $g=17.906\pm0.003$ & $g=16.510\pm0.004$ & $g=17.521\pm0.007$\\%PS1
		 & $r=16.952\pm0.005$ & 
         $r=17.707\pm0.003$ & 
         $r=16.447\pm0.002$ & 
         $r=17.323\pm0.002$\\
		 & $i=16.929\pm0.005$ & $i=17.646\pm0.003$ & $i=16.484\pm0.003$ & $i=17.274\pm0.004$\\
		 %SDSS& zmag=16.11 & zmag=15.17 & zmag=17.55 & zmag=16.38\\
		$E(B-V)$ & 0.18 & 0.15 & 0.00 & 0.10 \\
		Proper motion ($\rm{mas\ yr}^{-1}$) 
		&$\mu_\alpha\rm{cos}\delta$=$-12.138\pm0.090$
		& $\mu_\alpha\rm{cos}\delta$=$-14.817\pm0.098$
		& $\mu_\alpha\rm{cos}\delta$=$-11.819\pm0.054$
		& $\mu_\alpha\rm{cos}\delta$=$13.364\pm0.108$\\
		  & $\mu_\delta$=$-24.642\pm0.058$
		& $\mu_\delta$=$2.330\pm0.079$
		& $\mu_\delta$=$-7.881\pm0.060$ 
		& $\mu_\delta$=$-5.728\pm0.112$\\
		Heliocentric distance (kpc) 
		& $7.85\pm2.20$\tnote{d} 
		& $9.54\pm2.68$\tnote{d}
		& $6.61\pm1.85$\tnote{d} 
		& $8.68\pm2.43$\tnote{d}\\
		Radial velocity (km s$^{-1}$) 
		& $v_{\rm{los}}=-207.5\pm7.0$ 
		& $v_{\rm{los}}-318.6\pm2.7$ 
		& $v_{\rm{los}}=-417.1\pm6.7$
		& $v_{\rm{los}}=264.7\pm1.2$\\
		  & $v_{\rm{rf}}=-157.1\pm7.0$ 
		& $v_{\rm{rf}}=-187.2\pm2.6$ 
		& $v_{\rm{rf}}=-424.4\pm6.6$
		& $v_{\rm{rf}}=313.7\pm1.2$\\
		Total velocity (km s$^{-1}$) 
		& $806.1\pm276.6$ 
		& $627.4\pm168.5$ 
		& $511.0\pm59.2$  
		& $501.6\pm124.2$\\
		Galactocentric velocities (km s$^{-1}$) \tnote{b}
		 & $U = 90.9\pm70.6$& $U =-381.5\pm41.9$ & $U =-59.5\pm68.6$ & $U =-435.5\pm75.3$\\
		 & $V = -766.6\pm272.3$& $V =-225.2\pm86.4$ & $V =-93.5\pm97.1$& $V =-224.0\pm147.7$\\
		 & $W = -233.1\pm36.6$& $W =444.9\pm159.9$ & $W =-498.8\pm36.5$& $W =-108.0\pm19.9$\\
		Galactocentric distance $r$ (kpc) 
		& $4.77\pm0.58$
		& $6.29\pm1.27$
		& $12.47\pm1.48$
		& $15.30\pm2.21$\\
        Z (kpc) 
        & $4.18\pm1.16$
        & $4.29\pm1.20$
        & $6.06\pm1.66$
        & $-6.06\pm1.70$\\
		Spectral type \tnote{c} & A0III & A9V? & A0III & A0III\\
		$(BP-RP)_0$ & 0.34& 0.40& 0.40&0.46\\
		Absolute magnitude 
		& $M_G=2.01\pm0.60$\tnote{d} 
		& $M_G=2.44\pm0.60$\tnote{d}
		& $M_G=2.30\pm0.60$\tnote{d} 
		& $M_G=2.38\pm0.60$\tnote{d}\\
		\hline
	\end{tabular}
    \begin{tablenotes}
    \centering
        \item[a]The $G$ magnitudes are taken form the {\it Gaia} DR3 catalog, and the $g$/$r$/$i$ magnitudes are taken from the PanSTARRS1 catalog. 
        \item[b]This is the three-dimensional velocity components relative to the galactic center.
        \item[c]Given by the MKCLASS code.
        \item[d]From the line index - absolute magnitude relation of A-type stars.
    \end{tablenotes}
    \end{threeparttable}
    }
\end{table*}

\begin{table*}
	%\footnotesize
	\scriptsize
	\centering
	\caption{Properties of the new unbound HVS candidates LAMOST-HVS9-12}
	\label{tab:hvs9}
 \makebox[\textwidth][c]{
    \begin{threeparttable}
	\begin{tabular}{lcccc} % four columns, alignment for each
		\hline
		\hline
		& LAMOST-HVS9 & LAMOST-HVS10 & LAMOST-HVS11 & LAMOST-HVS12 \\
		\hline
		LAMOST designation 
		& J125538.89+352338.8 
		& J015725.24+323539.3 
		& J101018.82+302028.1 
		& J014106.87+163157.3\\
		Position ($\alpha^{\circ}, \delta^{\circ}$) (ICRS(Epoch  J2016.0)) 
		& (193.912015,35.394107) 
		& (29.355280,32.594310) 
		& (152.578431,30.341158) 
		& (25.278694,16.532547)\\
		($l^{\circ}, b^{\circ}$) 
		& (116.987838,81.685488) 
		& (138.689751,-28.250432) 
		& (197.946490,54.716176) 
		& (139.795126,-44.706447) \\
		Magnitudes
		& $G=17.428\pm0.003$ & $G=16.329\pm0.003$ & $G=16.338\pm0.003$ & $G=16.980\pm0.003$\\
		& $g=17.423\pm0.008$ & $g=16.354\pm0.003$ & $g=16.192\pm0.003$ & $g=16.885\pm0.007$\\%PS1
		& $r=17.501\pm0.003$ & $r=16.402\pm0.002$ & $r=16.533\pm0.004$ & $r=17.124\pm0.003$\\
		& $i=17.617\pm0.004$ & $i=16.477\pm0.004$ & $i=16.794\pm0.008$ & $i=17.355\pm0.003$\\
		$E(B-V)$ & 0.00 & 0.04 & 0.00 & 0.13 \\
		Proper motion ($\rm{mas\ yr}^{-1}$) 
		& $\mu_\alpha\rm{cos}\delta$=$-7.222\pm0.046$
		& $\mu_\alpha\rm{cos}\delta$=$14.859\pm0.070$
		& $\mu_\alpha\rm{cos}\delta$=$-0.891\pm0.061$
		& $\mu_\alpha\rm{cos}\delta$=$1.461\pm0.117$\\
		& $\mu_\delta$=$1.071\pm0.056$
		& $\mu_\delta$=$-7.980\pm0.071$
		& $\mu_\delta$=$-0.071\pm0.054$ 
		& $\mu_\delta$=$-1.564\pm0.093$\\
		Heliocentric distance (kpc) 
		& $14.16\pm3.94$\tnote{a} 
		& $11.20\pm0.82$\tnote{b}
		& $27.29\pm3.75$\tnote{c} 
		& $37.28\pm3.42$\tnote{d}\\
		%\footnote{geometric distance \textit{rgeo}}.
		Radial velocity (km s$^{-1}$) 
		& $v_{\rm{los}}=280.3\pm3.4$ 
		& $v_{\rm{los}}=169.8\pm2.9$ 
		& $v_{\rm{los}}=414.6\pm1.4$
		& $v_{\rm{los}}=384.3\pm4.0$\\
		& $v_{\rm{rf}}=319.7\pm3.4$ 
		& $v_{\rm{rf}}=305.0\pm2.8$ 
		& $v_{\rm{rf}}=369.5\pm1.4$
		& $v_{\rm{rf}}=488.3\pm3.9$\\
		Total velocity (km s$^{-1}$) 
		& $535.6\pm93.6$ 
		& $758.0\pm61.3$ 
		& $437.0\pm4.2$  
		& $508.7\pm12.8$\\
		Galactocentric velocities (km s$^{-1}$) 
		& $U =-449.9\pm122.0$& $U =-615.5\pm37.6$ & $U =-305.8\pm13.9$& $U =-299.1\pm19.1$\\
		& $V =68.8\pm57.3$& $V =-375.4\pm55.2$ & $V =138.6\pm7.9$& $V =80.5\pm34.6$\\
		& $W =282.5\pm3.5$& $W =-229.9\pm33.2$ & $W =279.6\pm10.2$& $W =-401.9\pm17.6$\\
		Galactocentric distance $r$ (kpc) 
		& $16.74\pm3.49$
		& $17.67\pm0.76$
		& $32.54\pm3.67$
		& $42.30\pm3.37$\\
		Z (kpc) 
		& $13.99\pm3.98$
		& $-5.27\pm0.38$
		& $22.39\pm3.07$
		& $-26.20\pm2.39$\\
		Spectral type & A1IV & A0V & B5V & B3V\\
		$(BP-RP)_0$ & 0.24& 0.25& $-0.21$&$-0.21$\\
		Absolute magnitude 
		& $M_G=1.68\pm0.60$\tnote{a} 
		& $M_G=1.00\pm0.16$\tnote{b}
		& $M_G=-0.84\pm0.30$\tnote{c} 
		& $M_G=-1.19\pm0.20$\tnote{d}\\
		\hline
	\end{tabular}
     \begin{tablenotes}
     \centering
    \item[a]From the line index - absolute magnitude relation of A-type stars.
    \item[b]From the theoretical absolute magnitude of A0V stars.
    \item[c]From the theoretical absolute magnitude of B5V stars.
    \item[d]From the theoretical absolute magnitude of B3V stars.
    \end{tablenotes}
    \end{threeparttable}
    }
\end{table*}

\begin{table*}
	%\footnotesize
	\scriptsize
	\centering
	\caption{Properties of the new unbound HVS candidates LAMOST-HVS13 and LAMOST-HVS14}
	\label{tab:hvs13}
 \makebox[\textwidth][c]{
    \begin{threeparttable}
	\begin{tabular}{lcccc} % four columns, alignment for each
		\hline
		\hline
		& LAMOST-HVS13 & LAMOST-HVS14 \\
		\hline
		LAMOST designation 
		& J212520.96+190802.1 
		& J015022.75+483631.1 \\
		Position ($\alpha^{\circ}, \delta^{\circ}$) (ICRS(Epoch  J2016.0)) 
		& (321.337282,19.133907) 
		& (27.594853,48.608665) \\
		($l^{\circ}, b^{\circ}$) 
		& (70.051942,-21.944436) 
		& (132.876072,-13.121732) \\
		Magnitudes
		& $G=16.245\pm0.003$ & $G=16.376\pm0.003$\\
		& $g=16.371\pm0.002$ & $g=16.358\pm0.003$\\%PS1
		& $r=16.305\pm0.003$ & $r=16.491\pm0.004$\\
		& $i=16.302\pm0.002$ & $i=16.643\pm0.004$\\
		$E(B-V)$ & 0.07 & 0.14\\
		Proper motion ($\rm{mas\ yr}^{-1}$) 
		& $\mu_\alpha\rm{cos}\delta$=$-8.535\pm0.057$
		& $\mu_\alpha\rm{cos}\delta$=$0.350\pm0.055$\\
		& $\mu_\delta$=$-7.985\pm0.037$
		& $\mu_\delta$=$-0.654\pm0.046$\\
		Heliocentric distance (kpc) 
		& $15.02\pm3.65$\tnote{a} 
		& $16.33\pm4.00$\tnote{b}\\
		%\footnote{geometric distance \textit{rgeo}}.
		Radial velocity (km s$^{-1}$) 
		& $v_{\rm{los}}=-398.3\pm9.4$ 
		& $v_{\rm{los}}=254.7\pm16.6$ \\
		& $v_{\rm{rf}}=-177.5\pm9.4$ 
		& $v_{\rm{rf}}=425.7\pm16.5$ \\
		Total velocity (km s$^{-1}$) 
		& $796.7\pm191.7$ 
		& $438.0\pm16.1$ \\
		Galactocentric velocities (km s$^{-1}$) 
		& $U =671.6\pm189.5$& $U =-178.4\pm12.5$\\
		& $V =-360.7\pm60.9$& $V =389.0\pm14.7$\\
		& $W =231.4\pm18.6$& $W =-91.7\pm11.3$\\
		Galactocentric distance $r$ (kpc) 
		& $14.63\pm3.04$
		& $22.52\pm3.86$\\
		Z (kpc) 
		& $-5.60\pm1.35$
		& $-3.65\pm0.90$\\
		Spectral type & B9III & B8V\\
		$(BP-RP)_0$ & 0.40& 0.25\\
		Absolute magnitude 
		& $M_G=0.515\pm0.525$\tnote{a} 
		& $M_G=-0.01\pm0.53$\tnote{b}\\
		\hline
	\end{tabular}
     \begin{tablenotes}
     \centering
    \item[a]From the best-fitted B9III star HD175640.
    \item[b]From the theoretical absolute magnitude of B8V stars.
    \end{tablenotes}
    \end{threeparttable}
    }
\end{table*}

\section{Summary}\label{sec:conclusions}
In this study, we conducted a systematic search for blue HVSs using the LAMOST DR10 LRS database. Our search resulted in a final sample of 125 high-velocity candidates, each with $\left| v_{\rm{rf} } \right| > 300$ km s$^{-1}$ or $v_{\rm{tot}} > 440$ km s$^{-1}$. This sample encompasses both previously discovered LAMOST HVSs and two additional known HVSs: SDSS J081828.07+570922.1 and HIP 60350. 

Significantly, we report the discovery of ten new unbound HVS candidates, designated as LAMOST-HVS5 to LAMOST-HVS14. Unlike earlier LAMOST searches that predominantly identified B-type stars, our current sample includes a notable proportion of A-type stars. Detailed kinematic analysis indicates that six of these candidates likely originated near the GC, whereas LAMOST-HVS5, HVS7, HVS13, and HVS14 appear unlikely to have originated from the GC. These findings are corroborated by our backward orbit integration analyses.

Although no hyper-velocity white dwarfs or subdwarfs were identified in this search, future studies focusing on non-DA type white dwarfs hold promise for uncovering new hyper-velocity WD and subdwarf stars.

A critical challenge in determining the orbits of these HVS candidates lies in accurate distance estimation, especially for stars without reliable parallax measurements. More precise distance calibrations, combined with improved proper motion data from {\it Gaia} DR4, will enhance our ability to constrain the origins of these candidates through orbit integration analysis.

Additionally, given the limited observation epochs for our candidates, binary systems with apparent velocity variations may introduce contamination into our sample, potentially leading to misclassification as HVS candidates. Follow-up spectroscopic observations will be essential to eliminate such cases, while medium- or high-resolution spectroscopy will provide detailed elemental abundances, offering deeper insights into the origins and characteristics of these stars.

\section*{Acknowledgements}

Y.H. acknowledges the supported from the National Science Foundation of China (NSFC Grant No. 12422303, 1209004) and the National Key R\&D Programme of China (Grant No. 2019YFA0405503).

This work has made use of data from the European Space Agency (ESA) mission
Gaia (\url{https://www.cosmos.esa.int/gaia}), processed by the {\it Gaia}
Data Processing and Analysis Consortium (DPAC,
\url{https://www.cosmos.esa.int/web/gaia/dpac/consortium}). Funding for the DPAC
has been provided by national institutions, in particular the institutions
participating in the {\it Gaia} Multilateral Agreement.

LAMOST is a National Major Scientific Project built by the Chinese Academy of Sciences, which has been provided by the National Development and Reform Commission.

\appendix

\section{Appendix information} \label{sec:appendix}

To derive the Balmer line index - $M_{G}$ relation for A-type stars, we utilize the LAMOST LRS Line-Index Catalog of A-Type Stars\footnote{\url{https://www.lamost.org/dr10/v1.0/catalogue}}. In order to achieve a more precise relation, we first define a rigorously selected sample within the catalog consisting of 680,989 spectra of 514,734 objects. 

\begin{figure}
	\begin{centering}
		\includegraphics[width=1.\columnwidth]{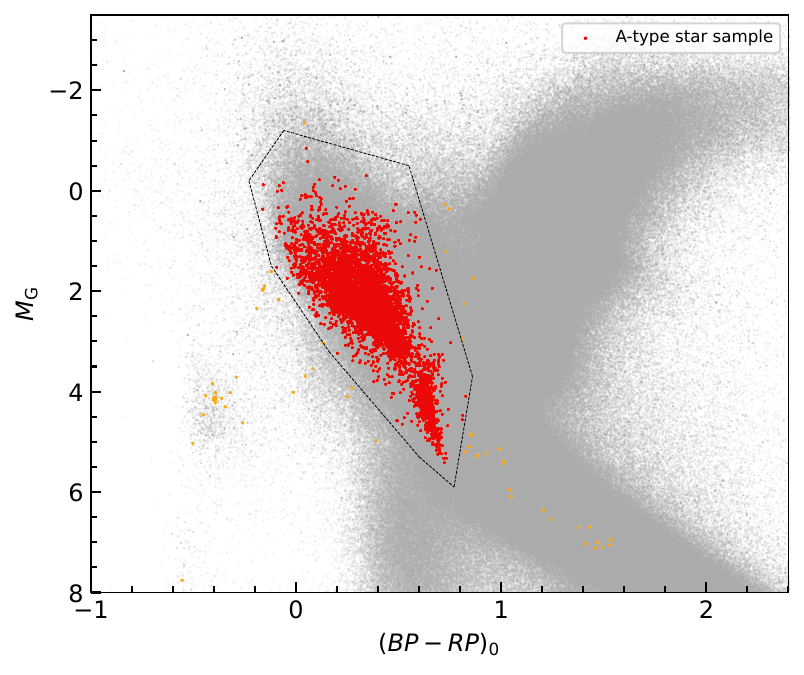}
		\caption{The A-type star sample consisting of 6,453 spectra from 5,190 objects defined to derive the Balmer line index - $M_{G}$ relation is represented as red circles in the {\it Gaia} CMD. Some hotsubdwarfs, low-temperature stars and other outliers represent in orange circles are excluded by the dotted lines to make the final sample more concentrated. All LAMOST DR10 LRS sources are plotted as grey points in the background for comparison.}
		\label{fig:as}
	\end{centering}
\end{figure}

To grantee accurate $M_{G}$ values, we impose quality control on the precision of parallax distances and extinction. We use the criteria of $rgeo < 3500$ pc to avoid large distance uncertainties, $RUWE < 1.4$ and $RPlx > 25$ in the {\it Gaia} DR3 data to ensure the distance accuracy. Additionally, to limit to low-extinction objects, we require the $E(B-V) < 0.7$ and Galactic latitude $\left| b \right| > 15$ degrees. The quality of the LAMOST data was further constrained to ensure more accurate line index measurements. We restrict the LAMOST redshift $\left| z \right|<0.001$ to exclude extragalactic sources and $fibermask==0$ to exclude bad spectra. The SNR cut $snru > 10$ and $snrg > 10$ and $snrr > 10$ and $snrg < 999$ is used to eliminate low-SNR spectra and avoid potentially problematic high-SNR spectra. The limiting magnitude of the LAMOST LRS is less than about 17.5, so the impact of {\it Gaia} photometric errors on $M_{G}$ is negligible. After applying these criteria, 6,524 spectra of 5,253 objects are remained. These objects are represented by orange and red circles in the {\it Gaia} CMD depicted in Figure~\ref{fig:as}. We then exclude objects lying outside of the black dotted region in the {\it Gaia} CMD, which are hot subdwarf stars, low-temperature objects and outliers. This results in a final sample of 6,453 spectra from 5,190 objects (red circles in Figure~\ref{fig:as}). They are used to derive the Balmer line index - $M_{G}$ relation.

\begin{figure}
	\begin{centering}
		\includegraphics[width=1.\columnwidth]{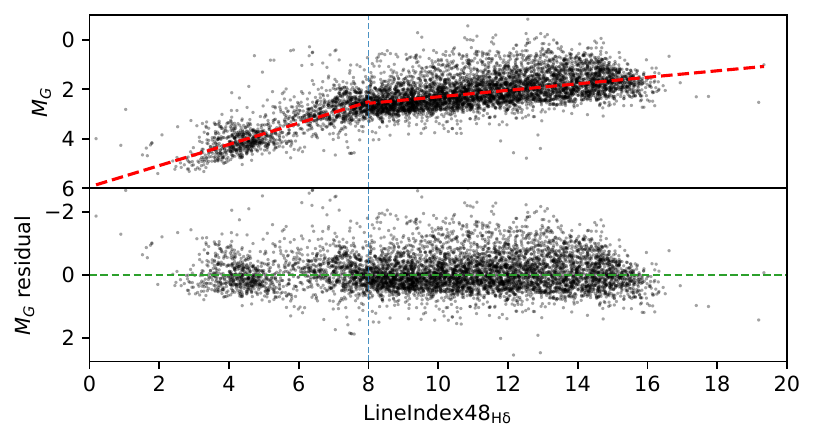}
		\caption{Upper panel: The final A-type star sample consisting of 6,453 spectra (black circles) give the relation between the H$\delta$ line index and $M_{G}$. The red dashed lines show the two-segment linear fitting result, and the blue dashed line represent the segment point at line index $= 8 $ \AA. Lower panel: The fitting residuals of $M_{G}$ are represented as black circles. The green dashed line represent $M_{G} = 0$ and the blue dashed line represent the fitting segment point at line index $= 8 $ \AA.}
		\label{fig:lineindex}
	\end{centering}
\end{figure}

We investigate the line index - $M_{G}$ relation using different Balmer lines and different band widths provided in the line index catalog. A two-segment linear function with the segment point at 8 \AA\ is employed to fit the relations. We find that the H$\delta$ line index with a 48 \AA\ band width make the relation retain well in the whole line index range and yield the smallest scatter in the fitting residuals. The fitting result is presented in Figure~\ref{fig:lineindex}. The distribution of the $M_{G}$ residual deviates slightly from a Gaussian distribution, with a 16th percentile value of $-0.60$, an 84th percentile value of 0.41, a median value of 0.00, and a standard deviation of 0.57. We adopt the value of 0.60 as the 1-$\sigma$ uncertainty for the prediction of $M_{G}$ from this relation.

\bibliography{bibliography}{}
\bibliographystyle{aasjournal}



\end{document}